\documentclass[12pt]{article}
\usepackage{graphicx,amssymb,amsmath}

\newcommand{\be}{\begin{equation}}
\newcommand{\ee}{\end{equation}}
\newcommand{\bea}{\begin{eqnarray}}
\newcommand{\eea}{\end{eqnarray}}
\newcommand{\beas}{\begin{eqnarray*}}
\newcommand{\eeas}{\end{eqnarray*}}

\newcommand{\hw}{\hat{\omega}}
\newcommand{\hk}{\hat{k}}

\begin{document}
\begin{titlepage}

\begin{center}

{\Large Asymmetric interiors for small black holes}

\vspace{8mm}

\renewcommand\thefootnote{\mbox{$\fnsymbol{footnote}$}}
Daniel Kabat${}^{1}$\footnote{daniel.kabat@lehman.cuny.edu},
Gilad Lifschytz${}^{2}$\footnote{giladl@research.haifa.ac.il}

\vspace{4mm}

${}^1${\small \sl Department of Physics and Astronomy} \\
{\small \sl Lehman College, City University of New York, Bronx NY 10468, USA}

\vspace{2mm}

${}^2${\small \sl Department of Mathematics} \\
{\small \sl Faculty of Natural Science, University of Haifa, Haifa 31905, Israel}

\end{center}

\vspace{8mm}

\noindent
We develop the representation of infalling observers and bulk fields in the CFT as a way to understand the black hole
interior in AdS.  We first discuss properties of CFT states which are dual to black holes.  We then show that in the
presence of a Killing horizon bulk fields can be decomposed into pieces we call ingoing and
outgoing.  The ingoing field admits a simple operator representation in the CFT, even inside a small black hole
at late times, which leads to a simple CFT description of infalling geodesics.  This means classical infalling
observers will experience the classical geometry in the interior.
The outgoing piece of the field is more subtle.  In an eternal two-sided geometry it
can be represented as an operator on the left CFT.  In a stable one-sided geometry it can be described using
entanglement via the PR
construction.  But in an evaporating black hole trans-horizon entanglement breaks down at the Page time, which means that for
old black holes the PR construction fails and the outgoing field does not see local geometry.  This picture of the interior allows the CFT to reconcile unitary Hawking
evaporation with the classical experience of infalling observers.

\end{titlepage}
\setcounter{footnote}{0}
\renewcommand\thefootnote{\mbox{\arabic{footnote}}}

\section{Introduction\label{sect:intro}}

\begin{figure}[!b]
\center{\includegraphics[width=9cm]{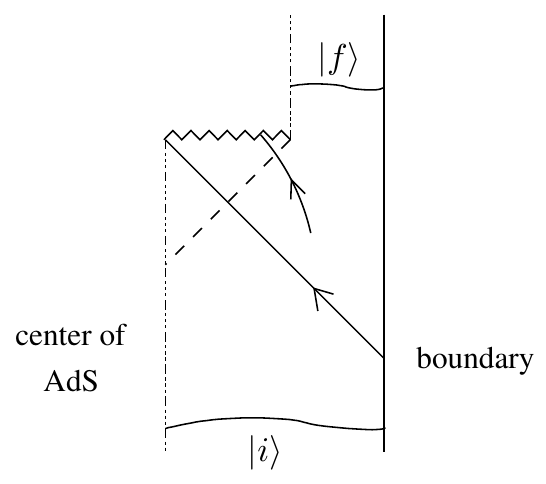}}
\caption{An evaporating AdS-Schwarzschild black hole, formed by a null shell sent in from the boundary.  The semiclassical geometry suggests that an infalling object will
hit the singularity and be lost to the outside world, while unitary time evolution requires the pure states $\vert i \rangle$ and $\vert f \rangle$ to be related by a unitary transformation.
\label{fig:evaporate}}
\end{figure}

Black holes provide an ideal theoretical laboratory for testing attempts to reconcile gravity with quantum mechanics.  There is a basic tension between the semiclassical geometry
thought to describe an evaporating black hole, shown in Fig.\ \ref{fig:evaporate}, and the requirement of unitary time evolution.
For a survey including recent developments see \cite{Harlow:2014yka}.

In this paper we assume that AdS/CFT provides a complete description of quantum gravity in asymptotically
AdS space.  This guarantees unitary time evolution for the underlying CFT degrees of freedom but leads one
to question the meaning of space-time inside the horizon. We probe this region by attempting to represent local bulk fields in the black hole interior in terms of the CFT.  For bulk points outside the horizon the
representation of a bulk field in terms of the CFT is well-understood: in the $1/N$ expansion
one can define CFT operators which mimic local bulk
fields when inserted in correlation functions.  This has been developed for free scalar fields \cite{Balasubramanian:1998sn,Banks:1998dd,Bena:1999jv,Hamilton:2005ju,Hamilton:2006az,Hamilton:2006fh}
and free fields with spin \cite{Heemskerk:2012np,Kabat:2012hp,Sarkar:2014dma}, and in empty AdS follows from representation theory \cite{Dobrev:1998md,Aizawa:2014yqa}.  Perturbative $1/N$ corrections
have been studied in \cite{Kabat:2011rz,Heemskerk:2012mn,Kabat:2012av,Kabat:2013wga,Kabat:2015swa}.
For other approaches see \cite{Verlinde:2015qfa,Miyaji:2015fia,Nakayama:2015mva}.

But inside
a black hole with finite entropy (and a single asymptotic region) it seems
unlikely that bulk fields directly correspond to CFT operators \cite{Marolf:2013dba}.
Given this, does the space-time region inside the horizon have any meaning?  We will argue that it does, in the sense that
even for evaporating black holes the CFT accurately describes the geometry seen by an infalling classical observer.

To show this we start from the simple observation that in the presence of a horizon a bulk field can be decomposed into parts we call ingoing and outgoing. This decomposition is crucial because, as we will see, the CFT treats these two parts of the field very differently.  Outside the horizon both parts can be represented as operators in the CFT.  But inside the horizon only the ingoing part of the field has a straightforward representation as an operator
in the CFT.\footnote{By `straightforward' we mean an
operator representation that follows from solving a wave equation on a given background
geometry.}  The outgoing piece does not have such a representation.
Fortunately Papadodimas and Raju (PR) \cite{Papadodimas:2012aq,Papadodimas:2013jku,Papadodimas:2015jra}
proposed a different method for representing fields in the interior which can be used to express the outgoing field
in the CFT.
The PR construction depends on entanglement across the horizon.  Given the maximal pairwise entanglement
expected from supergravity, and the known representation of local operators outside the horizon, one can write CFT operators
which act as local fields on the entangled partners in the interior.

The PR construction can be applied to an evaporating black hole, but there is a subtlety.
Unitarity of the CFT and monogamy of entanglement imply that after the Page time outgoing Hawking particles
are entangled, not with the black hole interior, but rather with the distant early Hawking radiation.  Once pairwise
trans-horizon entanglement is lost the
PR construction cannot be used to write local operators in the interior.  It seems that after the Page time there is no
way to recover conventional space-time geometry in the interior from the CFT.  Indeed it's been argued that
a firewall forms at the Page time \cite{Almheiri:2012rt}.

Here we argue for a different outcome. In classical gravity one way to show that the interior exists is to note that
infalling geodesics do not stop at the horizon but rather continue all the way
to the singularity.  We wish to make a similar argument in AdS/CFT.  To do this we construct bulk wavepackets
which track infalling geodesics.  We show that such wavepackets can be constructed using only the infalling
part of the field, which is the part that can be represented in the CFT.  These wavepackets track geodesics
which cross the horizon and continue all the way to the singularity.  This is true even after
the Page time!

So evaporating black holes do have an interior space-time geometry, in the restricted sense that
the CFT can describe classical objects falling into the black hole.  Although classical infall may be
geometric, for the outgoing part of the field the notion of interior geometry breaks down at the Page time.
So the CFT predicts no drama for infalling observers, while simultaneously realizing the partial breakdown of
geometry that is required for unitary evaporation.\footnote{Non-local models which account for unitary evaporation
have been discussed in \cite{Giddings:2011ks,Giddings:2012gc,Giddings:2013noa}.}

In the rest of the paper we elaborate on these statements.  Since the bulk of the paper is somewhat lengthy,
we provide below an overview of our approach.  Up to this point we have emphasized evaporating black holes,
but from here on we allow for more general possibilities.

\bigskip\smallskip
\centerline{* \hspace{2.5cm} * \hspace{2.5cm} *}
\smallskip

We start in section \ref{sect:BHstates} by discussing properties of CFT states which are dual to black
holes.  In particular we discuss the formation and evaporation of a black hole in the context of the eigenstate thermalization hypothesis,
contrasting the behavior of stable and evaporating black holes.  Although the discussion in this section is not strictly necessary
for the rest of the paper, it provides an important context for what follows.

In section \ref{sect:modes} we make the simple observation that in the presence of a horizon a field can be decomposed into ``ingoing'' and ``outgoing'' modes.  This is just the well-known fact that in terms of a tortoise
coordinate $r_*$, in the near-horizon region modes have two possible behaviors.
\bea
&& {\rm ingoing:} \qquad \phi \sim e^{-i \omega (t + r_*)} \\
\nonumber
&& {\rm outgoing:} \qquad \phi \sim e^{-i \omega (t - r_*)}
\eea
Note that ingoing modes are smooth across the future horizon while outgoing modes are singular.\footnote{Near the past horizon of an eternal black hole the behavior is reversed: the ingoing modes are singular and the outgoing modes
are smooth.}  The reason for this behavior
is that we're diagonalizing a Killing vector which is null on the horizon.  

Perhaps not surprisingly, the
ingoing part of the field has a simple representation as an operator in the CFT, even for bulk points inside the horizon.  We work out examples of these CFT operators in section \ref{sect:AdS2smear} for AdS${}_2$ in Rindler coordinates and in section \ref{sect:AdS3smear} for
AdS${}_3$ and BTZ black holes.  The simplest case is a free massless scalar field in AdS${}_2$, for which the ingoing
and outgoing parts of the field can be represented in terms of an operator ${\cal O}$ in the CFT by
\bea
&&\phi_{\rm in}(t,r_*) = {1 \over 2 R^2} \int_{t + r_*}^\infty dt' \, {\cal O}(t') \\[5pt]
\nonumber
&&\phi_{\rm out}(t,r_*) = - {1 \over 2 R^2} \int_{t - r_*}^\infty dt' \, {\cal O}(t')
\eea
Outside the horizon both $\phi_{\rm in}$ and $\phi_{\rm out}$ are well-defined and one recovers the full bulk
field from the combination $\phi_{\rm in} + \phi_{\rm out}$.  But as a CFT operator $\phi_{\rm in}$
smoothly extends across the future horizon into the interior.

To understand the significance of decomposing the field in this way, in section \ref{sect:near} we study the behavior of the ingoing and outgoing fields in the near-horizon region.  The outgoing modes are rapidly oscillating near the future horizon, so by the Riemann-Lebesgue lemma (and a proper
treatment of zero modes), on the future horizon
$\phi_{\rm out}$ vanishes and $\phi_{\rm in}$ agrees with the full field.  This provides an interpretation of
the decomposition into ingoing and outgoing fields.  $\phi_{\rm in}$ is non-normalizeable, so it describes a CFT with
sources turned on that
send excitations in from the boundary.  These sources are adjusted so that the field takes on the correct value on the future horizon.

Building on these results, in section \ref{sect:wavepackets} we argue that
the ingoing part of the field, which has a simple representation in the CFT, is sufficient to describe a wavepacket falling through
the horizon to very good accuracy.  We show this explicitly for AdS${}_2$ in Rindler coordinates, by constructing
wavepackets using the WKB approximation and
showing that in the geometric optics limit, where the WKB approximation becomes exact, a
description of the wavepacket solely in terms of the ingoing part of the field becomes possible.  Moreover in the
geometric optics limit wavepackets move along geodesics, and in this sense we claim that the CFT encodes the
interior geometry of the black hole.  Thus we learn from the CFT that, to very good accuracy, a classical observer freely falls across the horizon and experiences no drama until reaching
the singularity.

In section \ref{sect:interior} we use these results to reconsider the meaning of the black hole interior.  We investigate three distinct cases.

\goodbreak
\noindent{\em Eternal black holes} \\
For an eternal black hole with two asymptotic regions there is no difficulty representing local bulk fields in the interior, provided one
considers operators which act on both copies of the CFT.  The field in the interior can be written
as a superposition of an infalling field from the left and
an infalling field from the right.  In this sense an eternal black hole has a conventional internal geometry, with local bulk fields that can
be expressed in terms of CFT operators.

\goodbreak
\noindent{\em Stable black holes formed from collapse} \\
This differs from the eternal case in that there is only a single asymptotic region.  As discussed above, in the interior it is straightforward to
represent the ingoing part of the field as an operator in the CFT.  The outgoing part of the field does not have a conventional operator representation
in the CFT.  But it can be represented as a state-dependent
operator, using entanglement across the horizon and following the construction of Papadodimas and Raju.
In this sense a
stable black hole has conventional internal geometry, with however a hybrid description in the CFT:
the ingoing part of a field can be expressed as a conventional CFT operator while the
outgoing part can only be accessed using entanglement.

\goodbreak
\noindent{\em Unstable black holes} \\
Sufficiently small black holes in AdS are unstable and will eventually evaporate, just like
black holes in flat space.  As discussed above the ingoing part of the field has a straightforward operator representation in the CFT and experiences the classical geometry.  For the outgoing part one can apply the PR construction.  However the PR construction
relies on pairwise maximal entanglement of supergravity excitations across the horizon to describe local operators in the
interior.  So at first PR lets one describe a local outgoing field in the interior.  But around the Page time the pairwise trans-horizon entanglement required
for the construction of local mirror operators is lost \cite{Page:1993df,Page:1993wv,Braunstein:2009my}.  The most conservative assumption would seem to be, not that a firewall forms \cite{Almheiri:2012rt},
but that there are no local right-moving degrees of freedom in the interior of an old black hole.

Thus the CFT suggests a rather curious asymmetric
interior for an unstable black hole at late times.
Since the ingoing part of the field can describe an infalling classical
observer, while the outgoing part of the field describes Hawking particles and is responsible for the evaporation
process, this provides a mechanism for the CFT
to reconcile the semiclassical behavior of an infalling observer with the breakdown of geometry required for unitary Hawking evaporation.

\section{Black hole states in CFT \label{sect:BHstates}}

In this paper we will be concerned with the CFT description of black holes, including small black holes that are unstable and
eventually evaporate.  To provide a framework, in this section we discuss properties of CFT states and operators that are
expected to describe such black holes.  We consider stable and evaporating black holes in turn.  The calculations in the rest of the
paper do not depend on this section, so the impatient reader may skip ahead to section \ref{sect:modes}.

\subsection{CFT description of stable black holes}

We start by considering stable black holes which are formed from collapse.  Such black holes are dual to a pure initial state in the CFT
which is not thermal but evolves with time to a state that looks thermal for appropriately chosen observables.  Since the CFT is a closed
system this thermalization has to be understood without a heat bath.  This can be done using the Eigenstate Thermalization Hypothesis
(ETH) \cite{PhysRevA.43.2046,1994PhRvE..50..888S,1999JPhA...32.1163S}.  ETH explains how a closed system can evolve in a way that makes it look thermal after some time. ETH is conjectured to be correct for chaotic systems, which is consistent with the
connection between black hole horizons and chaos \cite{Shenker:2013pqa,Polchinski:2015cea}. ETH claims that in chaotic systems,
for eigenstates of the Hamiltonian $\vert\alpha\rangle$, $\vert \beta \rangle$ which are nearby in energy, there are operators ${\cal O}_{i}$ that obey
\be
\label{ETH}
\langle\beta\vert {\cal O}_{1} \cdots {\cal O}_{j} \vert \alpha \rangle = \delta_{\alpha \beta}{\cal A}_{1\cdots j}(E) +e^{-S(E)/2}f(E,\omega)R_{\alpha \beta}
\ee
Here $E=\frac{1}{2}(E_{\alpha}+E_{\beta})$, $\omega = E_{\alpha}-E_{\beta}$, and $S(E)$ is the microcanonical entropy.
${\cal A}_{1\cdots j}(E)$ and $f(E,\omega)$ are smooth functions of their arguments but $R_{\alpha\beta}$ is a numerical factor of order one which varies erratically with $\alpha$ and $\beta$.
The function ${\cal A}_{1\cdots j}(E)$ agrees with the microcanonical result for the correlation function up to very small corrections.
These properties ensure that for any given initial state, correlation functions at late times will be very close to the microcanonical
result.
Note that the smallness of the off-diagonal entries in ETH is such that no eigenstate is distinguished from any other. Since there
are $e^{S(E)}$ eigenstates and correlators in a generic state should be ${\cal O}(1)$ the ETH ansatz (\ref{ETH}) is the most democratic choice. This democracy is required
if we want all states (no matter what they are initially) to eventually thermalize.
Note that ETH is expected to apply to many but certainly not all operators.  In many-particle systems it is usually applied to operators
which measure single-particle properties.  In AdS/CFT we expect operators satisfying ETH to be
single-trace operators describing supergravity fields (and perhaps also some stringy excitations).

Now let's see how black hole formation is described by the CFT. We start with an initial state in the CFT 
\begin{equation}
\vert \psi \rangle = \sum_{\alpha} c_{\alpha} \vert \alpha \rangle
\end{equation}
where the sum runs over eigenstates of the Hamiltonian that have energy $E$ up to a small spread $\Delta E$.  The coefficients
$c_{\alpha}$ are chosen with care so that in the initial state correlators of supergravity operators are far from thermal. This is done by choosing the
initial phases of the $c_\alpha$ so that the off-diagonal entries in (\ref{ETH}), even though they are small, will add up to produce a result at least as large as the diagonal term. Under time evolution the $c_{\alpha}$ will get extra phases that destroy the original coherence.
So after some time  the contribution of the off-diagonal terms will be suppressed and correlation functions will look thermal to a good approximation. This is how thermalization is described using ETH.  One could say that thermalization is decoherence in the energy basis.  In AdS/CFT this process could describe the formation of a black hole.

In this picture it is easy to see why there appears to be information loss when a black hole is formed. All information about the state is contained the coefficients
$c_{\alpha}$. However after enough time passes that phase coherence has been lost, ETH gives
$\langle\psi\vert{\cal O}_{1}\cdots {\cal O}_{j}\vert\psi\rangle={\cal A}_{1\cdots j}(E)$
up to corrections of order $e^{-S(E)/2}$. This means correlators of supergravity operators are not sensitive to the values
$c_{\alpha}$. So in the supergravity approximation information about the microstate is lost.

The inability to distinguish which CFT state the system is in (using these operators) corresponds in the bulk to the
inability to distinguish horizon microstates using supergravity fields.  So we can equate the existence of a horizon with the validity
of the ETH ansatz.\footnote{We are claiming that ETH is a necessary but perhaps not sufficient condition for a horizon.}
In the gravity description the information is hidden behind the horizon, we will discuss later in what sense this is reflected in the CFT.

This description also makes it clear that early Hawking particles do not carry information \cite{Page:1993wv}. After the black hole is formed some early Hawking particles are produced which become the thermal atmosphere in AdS that the black hole is in equilibrium with. From the CFT perspective the production of these early Hawking particles is part of the thermalization process, so in fact the emission of these particles erases some of the information about the state. As another example, consider acting on the state after a time when it
looks thermal by annihilating some of the outside Hawking particles. This perturbed state is not generic (the number of particles
outside the black hole differs from the microcanonical average), but the black hole
will emit some particles and re-thermalize, loosing microstate information in the process.  This loss of information is due to the emission of Hawking particles.

The fact that in the semi-classical approximation one cannot determine the state does not of course mean that unitarity is lost. The CFT
state has undergone unitary evolution (in fact this is how ETH describes thermalization), but the set of operators that are available in the supergravity approximation only includes those whose correlation functions do not depend on the exact state (they are insensitive
to the values $c_{\alpha}$).
If we had access to the operator $\vert\alpha_i\rangle\langle\alpha_j\vert$ we could easily know the exact state. We can say that information about the state is encoded in non-geometric data.

\subsection{CFT description of evaporating black holes}

A black hole that forms from collapse and then evaporates has a time evolution which initially resembles that of a
stable black hole. One starts with a pure state that is far from equilibrium.  Under time evolution the system seems to thermalize and a black hole forms.  But the black hole is unstable and gradually evaporates.  The final state is well-described by a collection of
supergravity particles in AdS. What is the CFT description of such time evolution?
It must have a few remarkable properties. During an initial period of thermalization it must have some form of information loss (in the
supergravity approximation), but eventually all information must be present in supergravity correlation functions.

We have seen that ETH is related to many properties of the black hole, in particular to the initial collapse and formation of a horizon. But at late times ETH is not consistent with recovery of information, and in fact correlators in the thermal gas phase
are not compatible with ETH \cite{Barbon:2003aq}. It is also important to remember that  the state describing an evaporating black hole cannot be a typical state of the given energy. The entropy of a small black hole is less than the entropy of the thermal gas it is evaporating to (this is why it is evaporating), so states which go through a ``small black hole'' phase are not typical. 

We suggest the following description in terms of the CFT.
The initial state is a superposition of special states
which are only approximate eigenstates of the Hamiltonian.  These special states
span a small subspace of the full Hilbert space of the theory, and we assume that in these special states operators dual to supergravity fields obey ETH.
If we start with a superposition of these special states, for a while time evolution will not notice that they are only approximate
energy eigenstates, so initially a black hole forms and there is a horizon. However as time goes by since these special states are only approximate eigenstates of the Hamiltonian the system will leak out of the special subspace of the Hilbert space. This leakage is the evaporation of the black hole. Over sufficiently large times what matters is the exact eigenstates of the Hamiltonian, but these do not obey ETH. So at sufficiently late times information about the state can be deduced from supergravity correlators.

\section{Ingoing and outgoing modes\label{sect:modes}}

In this section we study the behavior of field modes near a Killing horizon and show that modes
of definite frequency can be characterized as either ingoing (smooth across the future horizon) or outgoing
(singular on the future horizon).  This behavior has been known since the early days \cite{PhysRevD.11.1404}.

For concreteness we focus on static metrics of the form
\be
\label{StaticMetric}
ds^2 = - f(r) dt^2 + {1 \over f(r)} dr^2 + r^2 ds^2_\perp
\ee
We assume that $f(r)$ vanishes, or equivalently that the Killing vector ${\partial \over \partial t}$ becomes null,
at some radius $r = r_0$.  Assuming a simple zero we have
\be
f(r) = {4 \pi \over \beta} (r - r_0) + {\cal O}\big((r - r_0)^2\big)
\ee
where $\beta$ is identified as the inverse temperature.  Some geometries which display this behavior are
\begin{itemize}
\item
Eternal AdS-Schwarzschild black holes, for which
\be
f(r) = {r^2 \over R^2} + 1 - {\omega_d M \over r^{d-2}}
\ee
Here $R$ is the AdS radius,
$M$ is the black hole mass, $\omega_d = {16 \pi G_N \over (d-1) {\rm vol}(S^{d-1})}$, and $ds^2_\perp$ is the metric on a round unit sphere $S^{d-1}$.
\item
AdS in Rindler coordinates, for which
\be
f(r) = {r^2 \over R^2} - 1
\ee
and $ds^2_\perp$ is the metric on the hyperbolic plane ${\cal H}^{d-1}$.
\item
The BTZ black hole, for which
\be
f(r) = {r^2 - r_0^2 \over R^2}
\ee
and the transverse space is a circle, $ds_\perp^2 = d\theta^2$ with $\theta \approx \theta + 2 \pi$.
\end{itemize}
We wish to study the wave equation $\big(\Box - m^2\big) \phi = 0$ in the geometry (\ref{StaticMetric}).
It's convenient to introduce a tortoise coordinate
\be
\label{tortoise}
r_* = \int^r {dr' \over f(r')}
\ee
so that
\be
ds^2 = f(r) \big(-dt^2 + dr_*^2\big) + r^2 ds_\perp^2
\ee
The integral (\ref{tortoise}) has a log divergence at the horizon, which means that
$r_* \rightarrow - \infty$ as $r \rightarrow r_0$.  For example for AdS-Rindler we have $r_0 = R$ and\footnote{With
asymptotic AdS boundary conditions it's convenient to set $r_* = - \int_r^\infty {dr' \over f(r')}$ so that
$r_* \rightarrow - \infty$ at the horizon and $r_* \rightarrow 0^-$ at the AdS boundary.}
\be
r_* = {R \over 2} \log {r - R \over r + R}
\ee
The wave equation can be solved by separating variables.
\be
\label{separate}
\phi(t,r_*,\Omega) = e^{-i \omega t} r^{1 - d \over 2} R(r_*) Y_k(\Omega)
\ee
Here $Y_k(\Omega)$ is a harmonic function of the transverse coordinates, $\Box_\perp Y_k = - k^2 Y_k$.
The ansatz (\ref{separate}) reduces the wave equation to a Schrodinger equation in an effective potential,
\be
\big[- \partial_{r_*}^2 + V(r_*)\big] R(r_*) = \omega^2 R(r_*)
\ee
where
\be
\label{Veff}
V(r_*) = f(r) \left[{k^2 \over r^2} + m^2 + {d-1 \over 2r} \, {df \over dr} + {(d-1)(d-3) \over 4r^2}
f\right]_{r = r(r_*)}
\ee
The important point is that, due to the prefactor $f(r)$, the potential vanishes at the horizon.
This just reflects the fact that the horizon is a surface of infinite redshift.  It means that in the near-horizon
region solutions to the wave equation have the form
\bea
\label{NearHorizon}
&& {\rm ingoing:} \qquad \phi \sim e^{-i \omega (t + r_*)} \\
\nonumber
&& {\rm outgoing:} \qquad \phi \sim e^{-i \omega (t - r_*)}
\eea
Approaching the future horizon $t \rightarrow + \infty$ and $r_* \rightarrow - \infty$, so the ingoing modes
are smooth while the outgoing modes oscillate rapidly.  Approaching the past horizon $t \rightarrow - \infty$
and $r_* \rightarrow - \infty$ so the behavior is reversed: the outgoing modes are smooth while
the ingoing modes oscillate rapidly.

\begin{figure}
\center{\includegraphics[width=9cm]{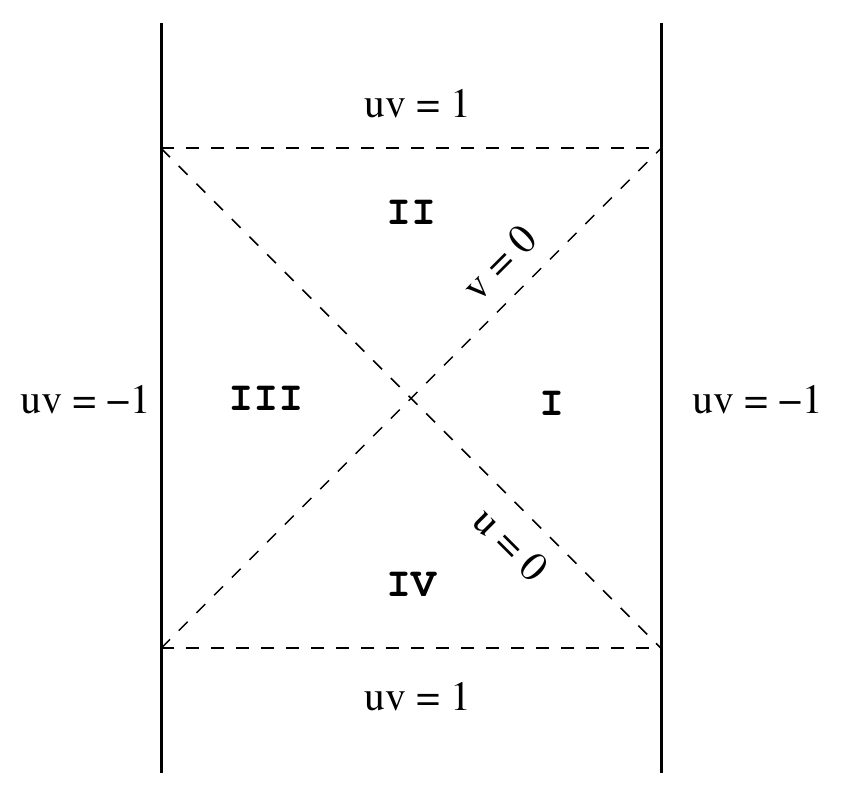}}
\caption{The Penrose diagram for AdS${}_2$ in Kruskal coordinates.
\label{fig:Penrose}}
\end{figure}

It's convenient to express this behavior in terms of Kruskal coordinates
\bea
&& u = e^{2 \pi (t + r_*) / \beta} \\
\nonumber
&& v = - e^{-2\pi (t - r_*) / \beta}
\eea
For asymptotic AdS space the boundary is at $uv = -1$, while the singularity is at $r = 0$ or equivalently
\cite{Klosch:1995qv,Fidkowski:2003nf}
\be
uv = \exp \left[ - {4 \pi \over \beta} {\rm PV} \int_0^\infty {dr' \over f(r')} \right]
\ee
For AdS${}_2$ the Penrose diagram is shown in Fig.\ \ref{fig:Penrose}.  In this case the $r = 0$ singularity
is at $uv = +1$ and is just a coordinate artifact.

In Kruskal coordinates the modes have the near-horizon behavior
\bea
\label{NearHorizon2}
&& {\rm ingoing:} \qquad \phi \sim u^{-i \omega \beta / 2 \pi} \\
\nonumber
&& {\rm outgoing:} \qquad \phi \sim (-v)^{i \omega \beta / 2 \pi}
\eea
This makes it clear that the ingoing modes are smooth across the future horizon while the outgoing modes
are singular.  Across the past horizon the behaviors are reversed: the outgoing
modes are smooth while the ingoing modes are singular.

Finally let us comment on the relevance of these results for the realistic case of a black
hole which is formed from collapse and subsequently evaporates.  Although our explicit calculations
are for static geometries, the near-horizon behavior (\ref{NearHorizon}), (\ref{NearHorizon2})
should hold quite universally in a short-wavelength approximation.  We expect it to be valid even for
evaporating black holes, to the extent that evaporation can be treated as an adiabatic process for the
modes of interest.

\section{Smearing functions in AdS${}_2$\label{sect:AdS2smear}}

We have seen that, in the presence of a horizon, a field can be decomposed into ingoing and outgoing
modes.  In this section we show how the ingoing and outgoing parts of the field can be represented
as operators in the CFT.  For simplicity we focus on AdS${}_2$ in Rindler coordinates, with metric
\be
ds^2 = - {r^2 - R^2 \over R^2} dt^2 + {R^2 \over r^2 - R^2} dr^2\,.
\ee
In the bulk it's more convenient to use Kruskal coordinates, defined by
\bea
\label{KruskalCoords}
&& u = \sqrt{r - R \over r + R} \, e^{t/R} \\
\nonumber
&& v = - \sqrt{r - R \over r + R} \, e^{-t/R}
\eea
so that
\be
ds^2 = - {4 R^2 du dv \over (1 + uv)^2}\,.
\ee
The Penrose diagram is shown in Fig.\ \ref{fig:Penrose}.

To get oriented consider a massless field in AdS${}_2$, dual to an operator of dimension $\Delta = 1$ in the
CFT.  The general solution to the free scalar wave equation is
\be
\phi(u,v) = \phi_{\rm in}(u) + \phi_{\rm out}(v)
\ee
That is, the familiar decomposition into left- and right-movers is the same as the decomposition into
ingoing modes (which depend on $u$) and outgoing modes (which depend on $v$).  We also need to
impose the boundary condition that the field vanishes as $uv \rightarrow -1$.  This requires
\be
\phi(u,v) = f(u) - f(-1/v)
\ee
or equivalently
\bea
&& \phi_{\rm in}(u) = f(u) \\
&& \phi_{\rm out}(v) = - f(-1/v)
\eea
In the rest of this section we show that, from the behavior near the right boundary, we can reconstruct $\phi_{\rm in}(u)$ for
$u > 0$ and $\phi_{\rm out}(v)$ for $v < 0$.  This will let us write CFT operators which
represent $\phi_{\rm in}$ in regions I and II of the Penrose diagram, and $\phi_{\rm out}$ in regions I and IV.

We develop this representation for the general case of a massive field in AdS${}_2$, dual to an operator of dimension
$\Delta = {1 \over 2} + \sqrt{{1 \over 4} + m^2 R^2}$ in the CFT.  The field can be expanded in a
complete set of normalizeable modes,
\be
\phi(u,v) = \int_{-\infty}^\infty d\omega \, a_\omega \phi_\omega(u,v)
\ee
where
\be
\label{AdS2phiomega}
\phi_\omega(u,v) = u^{-i \omega R} (1 + uv)^\Delta F(\Delta,\Delta-i\omega R, 2\Delta,1+uv)
\ee
These modes have definite frequency under Rindler time translation $t \rightarrow t + {\rm const}$.  They
can be decomposed into ingoing and outgoing pieces with the help of
some hypergeometric identities.
\bea
\label{decomposition}
&& \phi_\omega = \phi_\omega^{\rm in} + \phi_\omega^{\rm out} \\
\nonumber
&& \phi_\omega^{\rm in} = u^{-i \omega R} {\Gamma(2\Delta) \Gamma(i \omega R) \over
\Gamma(\Delta) \Gamma(\Delta + i \omega R)} F\big(\Delta,1-\Delta,1-i\omega R,{uv \over 1 + uv}\big) \\
\nonumber
&& \phi_\omega^{\rm out} = (-v)^{i\omega R} {\Gamma(2\Delta) \Gamma(-i \omega R) \over
\Gamma(\Delta) \Gamma(\Delta - i \omega R)} F\big(\Delta,1-\Delta,1+i\omega R,{uv \over 1 + uv}\big)
\eea
This decomposition illustrates the general near-horizon behavior discussed in section \ref{sect:modes}.

As $r \rightarrow \infty$ the field has normalizeable fall-off, $\phi(u,v) \sim r^{-\Delta} \phi_0(t)$, where $\phi_0(t)$
can be identified with an operator ${\cal O}_\Delta$ of dimension $\Delta$ in the CFT.  Sending $r \rightarrow \infty$ in the mode
expansion gives
\be
r^{-\Delta} \phi_0(t) = \int_{-\infty}^\infty d\omega \, a_\omega \Big({2 R \over r}\Big)^\Delta e^{-i\omega t}
\ee
which means
\be
\label{a_omega}
a_\omega = {1 \over (2 R)^\Delta} \int_{-\infty}^\infty {dt \over 2\pi} \, e^{i \omega t} \phi_0(t)
\ee
Plugging this back in the mode expansion lets us express the bulk field in terms of its near-boundary behavior,
\be
\phi(u,v) = \int_{-\infty}^\infty dt \, K(u,v \vert t) \, \phi_0(t)
\ee
where the smearing function $K$ is basically the Fourier transform of the mode functions.
\be
K = {1 \over (2 R)^\Delta} \int_{-\infty}^\infty {d\omega \over 2\pi} \, e^{i \omega t} \phi_\omega(u,v)
\ee
To express the ingoing and outgoing parts of the field in terms of the CFT we use the mode decomposition (\ref{decomposition}) to write 
\be
\phi(u,v) = \phi_{\rm in}(u,v) + \phi_{\rm out}(u,v)
\ee
where
\bea
\nonumber
&& \phi_{\rm in}(u,v) = \int_{-\infty}^\infty dt \, K_{\rm in}(u,v \vert t) \, \phi_0(t) \\
\label{Kin}
&& K_{\rm in} = {1 \over (2 R)^\Delta} \int_{-\infty}^\infty {d\omega \over 2\pi} \, e^{i \omega t} \phi_\omega^{\rm in}(u,v)
\eea
and
\bea
\nonumber
&& \phi_{\rm out}(u,v) = \int_{-\infty}^\infty dt \, K_{\rm out}(u,v \vert t) \, \phi_0(t) \\
\label{Kout}
&& K_{\rm out} = {1 \over (2 R)^\Delta} \int_{-\infty}^\infty {d\omega \over 2\pi} \, e^{i \omega t} \phi_\omega^{\rm out}(u,v)
\eea

When $\Delta$ is an integer the Fourier transforms in (\ref{Kin}), (\ref{Kout}) simplify since the modes reduce to
elementary functions with a finite number of poles.  We proceed to consider a few examples.

\bigskip
\noindent
{\em Massless field, $\Delta = 1$}

\noindent
In this case the normalizeable mode (\ref{AdS2phiomega}) reduces to
\be
\phi_\omega(u,v) = {1 \over i \omega R} \left[u^{-i \omega R} - (-v)^{i \omega R}\right]
\ee
so that
\bea
&& \phi_\omega^{\rm in} = {1 \over i \omega R} u^{-i \omega R} \\
\nonumber
&& \phi_\omega^{\rm out} = - {1 \over i \omega R} (-v)^{i \omega R}
\eea
The splitting of the zero mode into ingoing and outgoing pieces is ambiguous.  We resolve the ambiguity
with an $i \epsilon$ prescription, defining
\bea
\nonumber
K_{\rm in} & = & {1 \over 2 R^2} \int_{-\infty}^\infty {d\omega \over 2 \pi i} \, {1 \over \omega - i \epsilon}
e^{i \omega (t - R \log u)} \\
& = & {1 \over 2 R^2} \, \theta(t - R \log u)
\eea
and
\bea
\nonumber
K_{\rm out} & = & - {1 \over 2 R^2} \int_{-\infty}^\infty {d\omega \over 2 \pi i} \, {1 \over \omega - i \epsilon}
e^{i \omega (t + R \log (-v))} \\
& = & - {1 \over 2 R^2} \, \theta(t + R \log (-v))
\eea
Thus we can define CFT operators which mimic the ingoing and outgoing parts of the bulk field.
\bea
\label{DeltaEq1phiin}
&& \phi_{\rm in} = {1 \over 2 R^2} \int\limits_{R \log u}^\infty \!\! dt \, {\cal O}(t) \\
&& \phi_{\rm out} = - {1 \over 2 R^2} \int\limits_{R \log (-1/v)}^\infty \!\!\! dt \, {\cal O}(t)
\eea
For points in the right Rindler wedge note that $0 < u < -1/v$, so we recover the usual expression
for a massless bulk field \cite{Hamilton:2005ju}
\be
\phi = {1 \over 2 R^2} \int\limits_{R \log u}^{R \log (-1/v)} \!\!\! dt \, {\cal O}(t)
\ee
But note that the expression for $\phi_{\rm in}$ extends smoothly across the future horizon into region II
of the Penrose diagram, while $\phi_{\rm out}$ extends smoothly across the past horizon into region IV.

\bigskip
\noindent
{\em Massive field with $\Delta = 2$}

\noindent
To illustrate a more generic case we consider a massive field with $\Delta = 2$.  For $\Delta = 2$ the
Fourier transforms (\ref{Kin}), (\ref{Kout}) reduce to
\be
K_{\rm in} = {3 \over 2 R^2} \int_{-\infty}^\infty {d\omega \over 2\pi} e^{i \omega (t - R \log u)}
{1 \over i (\omega - i \epsilon) R (1 + i \omega R)} \left(1 - {2 u v \over (1 + uv)(1 - i \omega R)}\right)
\ee
and
\be
K_{\rm out} = {3 \over 2 R^2} \int_{-\infty}^\infty {d\omega \over 2\pi} e^{i \omega (t + R \log (-v))}
{1 \over - i (\omega - i \epsilon) R (1 - i \omega R)} \left(1 - {2 u v \over (1 + uv)(1 + i \omega R)}\right)
\ee
where we introduced an $i \epsilon$ prescription to handle the zero mode ambiguity.  The integrals are
straightforward and lead to
\bea
\nonumber
\phi_{\rm in} & = & - {3 \over 2 R^3} \int_{-\infty}^{R \log u} dt \, {1 \over 1 + uv} \, v e^{t/R} \, {\cal O}(t) \\
\label{DeltaEq2phiin}
& & + {3 \over 2 R^3} \int_{R \log u}^\infty dt \, {1 \over 1 + uv} \left(1 - uv - u e^{-t/R}\right) {\cal O}(t) \\ [10pt]
\nonumber
\phi_{\rm out} & = & {3 \over 2 R^3} \int_{-\infty}^{R \log (-1/v)} dt \, {1 \over 1 + uv} \, v e^{t/R} \, {\cal O}(t) \\
& & - {3 \over 2 R^3} \int_{R \log (-1/v)}^\infty dt \, {1 \over 1 + uv} \left(1 - uv - u e^{-t/R}\right) {\cal O}(t)
\eea
Again the combination $\phi_{\rm in} + \phi_{\rm out}$ is defined in the right Rindler wedge and matches
the usual expression for a bulk field \cite{Hamilton:2005ju}.  But $\phi_{\rm in}$ extends across the future horizon
into region II, while $\phi_{\rm out}$ extends across the past horizon into region IV.  Also note that, as a consequence
of our $i \epsilon$ prescription, the ingoing and outgoing smearing functions vanish exponentially as $t \rightarrow -\infty$.

\section{Smearing for AdS${}_3$ and BTZ black holes\label{sect:AdS3smear}}

In this section we extend the discussion of smearing functions to AdS${}_3$ and BTZ black holes.  Our goal is
to write down operators which represent the ingoing and outgoing parts of the field in terms of the CFT.

To treat AdS${}_3$ and BTZ in parallel we take the metric
\be
ds^2 = - {r^2 - r_0^2 \over R^2} dt^2 + {R^2 \over r^2 - r_0^2} dr^2 + r^2 d\theta^2 \qquad -\infty < \theta < \infty
\ee
This becomes AdS${}_3$ in Rindler coordinates when $r_0 = R$ and $\theta$ is non-compact.
It becomes a BTZ black hole when $\theta$ is periodically identified,
$\theta \approx \theta + 2 \pi$.

Consider a scalar field of mass $m$.  The field has an expansion in a complete set of modes
\be
\phi(t,r,\theta) = \int_{-\infty}^\infty d\omega \int_{-\infty}^\infty dk \, a_{\omega k} e^{-i \omega t} e^{i k \theta} \phi_{\omega k}(r)
\ee
where
\be
\phi_{\omega k}(r) = r^{-\Delta} \left({r^2 - r_0^2 \over r^2}\right)^{-i \hw / 2}
F\left({\Delta - i \hw - i \hk \over 2}, {\Delta - i \hw + i \hk \over 2}, \Delta,{r_0^2 \over r^2}\right)
\ee
and we've defined $\hw = \omega R^2 / r_0$, $\hk = k R / r_0$.  As $r \rightarrow \infty$ the field has normalizeable
fall-off, $\phi(t,r,\theta) \sim r^{-\Delta} {\cal O}_\Delta(t,\theta)$, where ${\cal O}_\Delta$ is an operator of dimension
$\Delta = 1 + \sqrt{1 + m^2 R^2}$ in the CFT.

In attempting to reconstruct $\phi$ from its near-boundary behavior one faces the problem of reconstructing
an evanescent wave \cite{Leichenauer:2013kaa,Rey:2014dpa}.  This can be done by complexifying the boundary \cite{Hamilton:2006az} or by regarding the
smearing function not as a function but as a distribution \cite{Morrison:2014jha}.  Here we will avoid these issues by working in a sector with fixed
spatial momentum $k$, so that all fields have a spatial dependence $e^{i k \theta}$ which we will suppress.  This approach was
also adopted in \cite{Papadodimas:2012aq}.  For AdS-Rindler $k$ is continuous while for BTZ $k \in {\mathbb Z}$.

Just as in the last section, for fixed spatial momentum $k$ we can reconstruct the bulk field via
\be
\phi_k(t,r) = \int dt' \, K_k(t,r \vert t') {\cal O}_{\Delta k}(t')
\ee
where the smearing function $K_k$ is a Fourier transform of the field modes.
\be
K_k(t,r \vert t') = \int_{-\infty}^\infty {d \omega \over 2 \pi} e^{-i \omega (t - t')} \phi_{\omega k}(r)
\ee

Now let's decompose the field into ingoing and outgoing pieces.  A hypergeometric transformation gives
\be
\phi_{\omega k} = \phi_{\omega k}^{\rm in} + \phi_{\omega k}^{\rm out}
\ee
where the in and out modes can be distinguished by their near-horizon ($r \rightarrow r_0$) behavior.
\bea
\label{PhiOmegaKIn}
&& \hspace{-2.1cm} \phi_{\omega k}^{\rm in} = r^{-\Delta} \left({r^2 - r_0^2 \over r^2}\right)^{- i \hw / 2}
{\Gamma(\Delta) \Gamma(i \hw) \over \Gamma(\Delta_{++}) \Gamma(\Delta_{+-})}
F\big(\Delta_{--},\Delta_{-+},1-i\hw,{r^2 - r_0^2 \over r^2}\big) \\
\nonumber
&& \hspace{-2cm} \phi_{\omega k}^{\rm out} = r^{-\Delta} \left({r^2 - r_0^2 \over r^2}\right)^{i\hw / 2}
{\Gamma(\Delta) \Gamma(-i \hw) \over \Gamma(\Delta_{--}) \Gamma(\Delta_{-+})}
F\big(\Delta_{++}, \Delta_{+-}, 1 + i \hw, {r^2 - r_0^2 \over r^2}\big)
\eea
Here $\Delta_{\pm\pm} = {1 \over 2} \big(\Delta \pm i \hw \pm i \hk\big)$.  In terms of the tortoise coordinate
\be
r_* = {R^2 \over 2 r_0} \log {r - r_0 \over r + r_0}
\ee
the near-horizon behavior is as expected: $\phi^{\rm in}_{\omega k} \sim e^{-i \omega r_*}$,
$\phi^{\rm out}_{\omega k} \sim e^{+i \omega r_*}$.

The ingoing and outgoing smearing functions $K_k^{\rm in}$, $K_k^{\rm out}$ are the Fourier transforms of these modes.
It's straightforward to evaluate the integrals but the results are not very enlightening.
For example, to evaluate $K_k^{\rm in}$, note that $\Gamma(i\hw)$ has simple poles at\footnote{The pole at $\omega = 0$ can be handled as in the previous section, with an $\omega \rightarrow \omega - i \epsilon$
prescription.}
\be
\omega = i n r_0 / R^2 \qquad n = 0,1,2,\ldots
\ee
while the hypergeometric function has simple poles at
\be
\omega = -i n r_0 / R^2 \qquad n = 1,2,3,\ldots
\ee
For large $\vert \omega \vert$ the mode $\phi_{\omega k}^{\rm in}$ behaves exponentially,\footnote{We're only keeping track of
the exponential dependence
on $\omega$.  To see this note that for the general static metric (\ref{StaticMetric}) the modes satisfy
${1 \over r^{d-1}} \partial_r r^{d-1} f(r) \partial_r \phi_{\omega k} + \left({\omega^2 \over f} - {k^2 \over r^2} - m^2\right) \phi_{\omega k}
= 0$.  For large $\omega$ the WKB approximation gives $\phi_{\omega k}^{\rm in}
\sim {\cal N}_\omega e^{-i \omega r_*}$.  By studying the $r \rightarrow r_0$ behavior of (\ref{PhiOmegaKIn}) one can show
that the normalization ${\cal N}_\omega$ introduces no additional exponential dependence on $\omega$.}
\be
\phi_{\omega k}^{\rm in} \sim e^{-i \omega r_*}
\ee
So for
\be
t' > t + r_*
\ee
we can close the contour in the upper half plane to find
\be
K_k^{\rm in}(t,r \vert t') = {r_0 \, \Gamma(\Delta) \over \! R^2} \sum_{n = 0}^\infty {(-1)^n \over n!} e^{-n r_0 (t' - t) / R^2}  f_{nk}(r)
\ee
Likewise for $t' < t + r_*$ we close in the lower half plane and have
\be
K_k^{\rm in}(t,r \vert t') = - {r_0 \Gamma(\Delta) \over R^2} \sum_{n = 1}^\infty {(-1)^n \over n!} e^{- n r_0 (t - t') / R^2} f_{nk}(r)
\ee
In these expressions we've defined
\be
f_{nk}(r) = {1 \over r^\Delta} \left({r^2 - r_0^2 \over r^2}\right)^{n/2} {F\big({\Delta + n + i \hk \over 2}, {\Delta + n - i \hk \over 2}, n+1, {r^2 - r_0^2 \over r^2}\big) \over
\Gamma\big({\Delta - n + i \hk \over 2}\big) \Gamma\big({\Delta - n - i \hk \over 2}\big)}
\ee
The outgoing smearing functions can be evaluated in the same way.  We find that for $t' > t - r_*$
\be
K_k^{\rm out}(t,r \vert t') = -{r_0 \, \Gamma(\Delta) \over \! R^2} \sum_{n = 0}^\infty {(-1)^n \over n!} e^{-n r_0 (t' - t) / R^2}  f_{nk}(r)
\ee
and for $t' < t - r_*$
\be
K_k^{\rm out}(t,r \vert t') = {r_0 \Gamma(\Delta) \over R^2} \sum_{n = 1}^\infty {(-1)^n \over n!} e^{- n r_0 (t - t') / R^2} f_{nk}(r)
\ee
Note that both $K_{\rm in}$ and $K_{\rm out}$ decay exponentially on the boundary in the far past, that is as $t' \rightarrow - \infty$.
Due to our $i \epsilon$ prescription they both approach constants in the far future, as $t' \rightarrow + \infty$.
Outside the horizon one can form the combination $K = K_{\rm in} + K_{\rm out}$ and use it to recover the full field $\phi$.
There's an amusing cancellation which makes $K$ non-zero only at spacelike separation, that is for $t + r_* < t' < t - r_*$.

\begin{figure}
\center{\hspace*{1cm} \includegraphics[width=8cm]{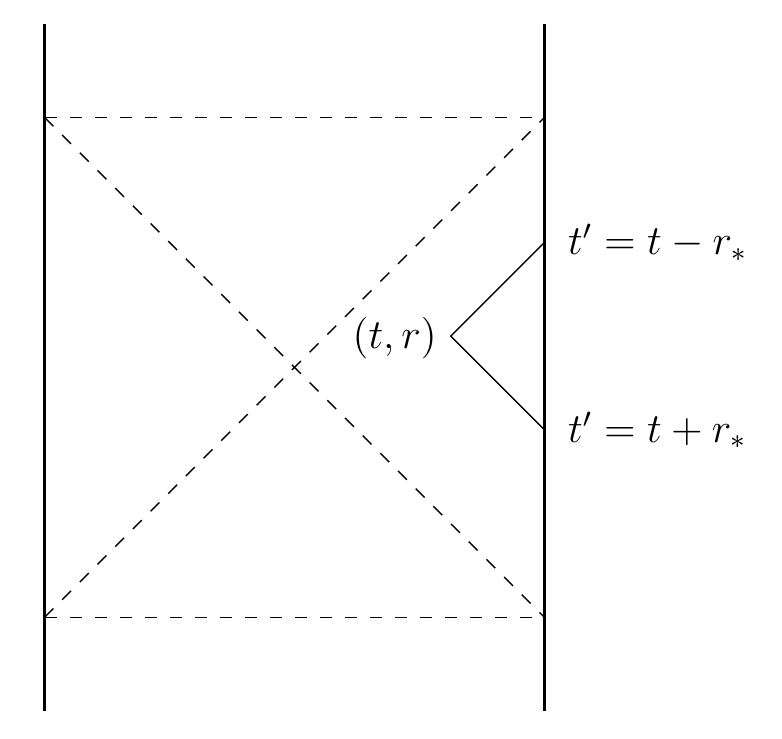}}
\caption{An AdS${}_2$ slice through AdS${}_3$.  The ingoing smearing function is non-analytic at $t' = t + r_*$ and the outgoing
smearing function is non-analytic at $t' = t - r*$.
\label{fig:lightcones}}
\end{figure}

Although these expressions are not very enlightening, there is an important lesson here.  The ingoing smearing function is non-analytic
at $t' = t + r_*$, which is exactly the time when a past-directed radial null geodesic from the bulk point would hit the boundary.
Likewise, as shown in Fig.\ \ref{fig:lightcones}, the outgoing smearing function is non-analytic when the future-directed radial null geodesic hits the boundary.  This behavior means there's no obstacle to continuing $K_k^{\rm in}$ across the future horizon to define an ingoing field in the
future interior.  Likewise there's no obstacle to continuing $K_k^{\rm out}$ across the past horizon.\footnote{One can
rewrite the smearing functions in Kruskal coordinates to make this a bit more manifest.}

\section{Near-horizon behavior\label{sect:near}}

In this section we study the behavior of the ingoing and outgoing parts of the field in the near-horizon region.  This leads to
an understanding of the ingoing field, as describing a CFT deformed by sources which are set up to
create the correct field profile on the future horizon.  It will also shed light on the interpretation of $\phi_{\rm in}$ in the interior
region, as providing a solution in the interior which satisfies certain boundary conditions on the horizon.

For simplicity we treat AdS${}_2$ in Kruskal coordinates.
In the right Rindler wedge a normalizeable bulk field has a mode expansion
\be
\phi(u,v) = \int_{-\infty}^\infty d\omega \, a_\omega \phi_\omega(u,v)
\ee
where $\phi_\omega$ is given in (\ref{AdS2phiomega}).  The field can be decomposed into ingoing and outgoing pieces,
\bea
&& \phi_{\rm in} = \int_{-\infty}^\infty d\omega \, a_\omega \phi_\omega^{\rm in} \\
\nonumber
&& \phi_{\rm out} = \int_{-\infty}^\infty d\omega \, a_\omega \phi_\omega^{\rm out}
\eea
where the ingoing and outgoing modes are given in (\ref{decomposition}).

Near the AdS boundary (where $uv \rightarrow -1$) the modes $\phi_\omega$ are normalizeable, with $\phi_\omega \sim (1 + uv)^\Delta$.
But the in and out modes are generically non-normalizeable, with $\phi_\omega^{\rm in,\,out} \sim (1 + uv)^{1-\Delta}$.  Clearly
$\phi_{\rm in}$ and $\phi_{\rm out}$ are bad approximations to the full field near the AdS boundary.  But we're
interested
in studying $\phi_{\rm in}$ inside the horizon, where the
near-boundary behavior doesn't matter, and where $\phi_{\rm in}$ provides a perfectly good solution to the equations of motion.  To complete the picture we'd like to understand how $\phi_{\rm in}$ and $\phi_{\rm out}$ behave near the horizon, since the horizon
provides a Cauchy surface for the interior.

It's straightforward to study the near-horizon behavior.  Let's start with $\phi_{\rm in}$, which
has the mode expansion
\be
\phi_{\rm in} = \int_{-\infty}^\infty d\omega \, a_\omega u^{-i \omega R} {\Gamma(2\Delta) \Gamma(i \omega R) \over
\Gamma(\Delta) \Gamma(\Delta + i \omega R)} F\big(\Delta,1-\Delta,1-i\omega R,{uv \over 1 + uv}\big)
\ee
The $\Gamma$ functions contribute poles at $\omega = i n / R$, $n = 0,1,2,\ldots$ while the hypergeometric function contributes
poles at $\omega = - i n / R$, $n = 1,2,3,\ldots$  Deforming the integration contour to pass below the pole at $\omega = 0$,\footnote{This matches the $i \epsilon$ prescription we introduced in section \ref{sect:AdS2smear}.} as
$u \rightarrow 0$ we can deform the contour upward to obtain the (presumably asymptotic) expansion\footnote{When
$\Delta$ is an integer the sum truncates and the hypergeometric function reduces to a finite polynomial.}
\be
\phi_{\rm in} = {2 \pi \Gamma(2\Delta) \over R \Gamma(\Delta)} \sum_{n = 0}^\infty {(-1)^n \over n! \, \Gamma(\Delta - n)}
F(\Delta, 1-\Delta, 1+n, {uv \over 1 + uv}) u^n a \vert_{\omega = i n / R}
\ee
where we've assumed that $a_\omega$ is an entire function.  Likewise as $v \rightarrow 0$ $\phi_{\rm out}$ has the expansion
\be
\phi_{\rm out} = {2 \pi \Gamma(2\Delta) \over R \Gamma(\Delta)} \sum_{n = 1}^\infty {(-1)^n \over n! \, \Gamma(\Delta - n)}
F(\Delta, 1-\Delta, 1+n, {uv \over 1 + uv}) (-v)^n a \vert_{\omega = - i n / R}
\ee
which follows from deforming the integration contour downward.

Note that $\phi_{\rm out}$ vanishes as $v \rightarrow 0$, which means that $\phi_{\rm in}$ must agree with the full field
on the right future horizon.
This gives a physical interpretation of $\phi_{\rm in}$.  Since the in and out fields
are non-normalizeable they cannot be identified with excited states in the CFT \cite{Balasubramanian:1998sn}.  Instead $\phi_{\rm in}$
describes a deformed CFT, with sources turned on to send excitations in from the right boundary.
The sources are
adjusted to reproduce the full field profile on the right future horizon.

Also note that, due to our $i \epsilon$ prescription, $\phi_{\rm in}$ has a zero mode contribution as $u \rightarrow 0$.
So on the left future horizon $\phi_{\rm in}$ doesn't quite vanish, instead it's given by the zero mode.
This leads to another perspective on $\phi_{\rm in}$.  The horizon provides a Cauchy surface for the interior, and since it's a null Cauchy surface the value of the field is sufficient initial data for the wave equation.
(In light-front coordinates the wave equation is first-order in time derivatives.)
So $\phi_{\rm in}$ is the unique solution in the interior which agrees with the full field on the right future horizon
and is given by the zero mode on the left future horizon.

Although our explicit calculations are for two-dimensional AdS-Rindler space, we expect that a similar discussion should
apply to an eternal AdS-Schwarzschild black hole.

\section{Infalling wavepackets\label{sect:wavepackets}}

In this section we show that the ingoing part of the field is capable of describing localized wavepackets that fall through the horizon and move along infalling geodesics.  Most of our analysis in this section will be classical, and by  ``wavepacket'' we will
mean a spatially-localized solution to the classical wave equation, although at the end we comment on the extension to
the quantum theory.  For simplicity we focus on wavepackets in AdS${}_2$, although the qualitative conclusions
should hold more generally.

We will be interested in wavepackets that provide a good semiclassical approximation to particle geodesics -- that is, in
the sort of wavepacket that can be used to describe a semiclassical observer falling into a black hole.  There is an important point of
principle here.  In the framework of field theory in curved space one often introduces the notion of an ``external observer'':
someone who can move along an arbitrary timelike trajectory, and who carries a particle detector (usually modeled as a quantum system with
discrete energy levels) that is coupled to the field at the position of the observer \cite{Unruh:1976db,Birrell:1982ix}.
In the framework of field theory in curved space it makes sense to introduce such an external observer,\footnote{This can be done in a systematic approximation, since
back-reaction is under control for observers that are light compared to the Planck scale.} but in the context of AdS/CFT one does not have this luxury.  Unless one modifies the CFT in some way, the only type of observer that is allowed is an ``internal observer'': an object that can be self-consistently described as an on-shell excitation of the available bulk degrees
of freedom.  In the leading large-$N$ limit, this means the only type of observer one can introduce is a free wavepacket falling into a
black hole.

\begin{figure}
\center{\hspace*{1cm} \includegraphics[width=8cm]{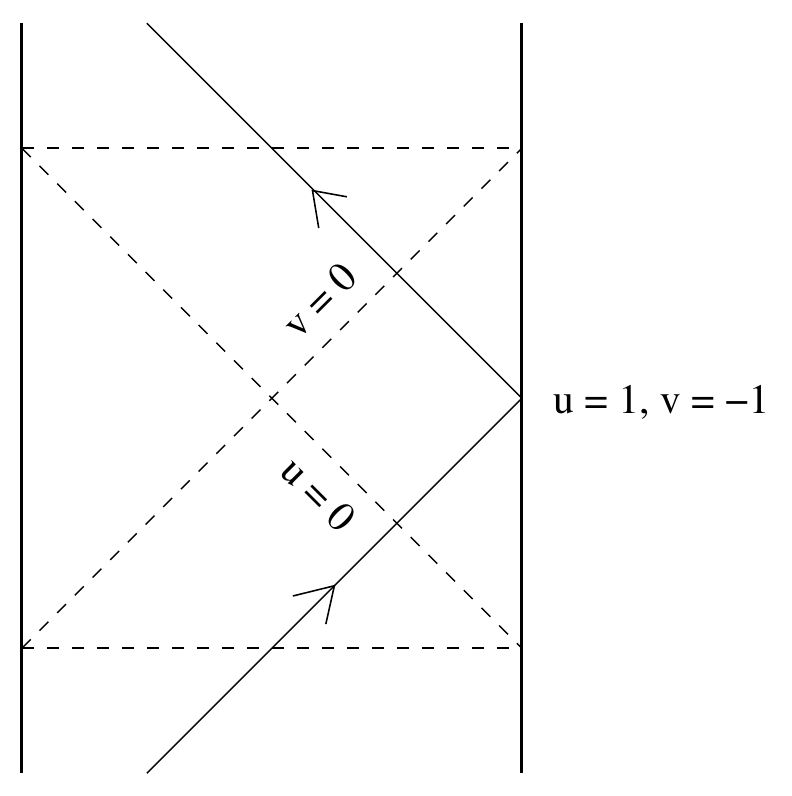}}
\caption{A null geodesic in AdS${}_2$ that bounces off the boundary at $u = 1$, $v = -1$.
\label{fig:AdS2geodesics}}
\end{figure}

To get oriented let's consider a massless field in AdS${}_2$, much as we did near
the beginning of section \ref{sect:AdS2smear}.  In this case particle geodesics are easy to describe.  As shown in Fig.\ \ref{fig:AdS2geodesics} they're null lines that bounce back
and forth between the two AdS boundaries.\footnote{In the two-dimensional Einstein static universe ${\mathbb R} \times S^1$ the bouncing geodesic lifts to a pair
of null lines that spiral around the cylinder in opposite directions.}  Wavepackets are equally easy to describe.  With Dirichlet
boundary conditions the general solution to the equations of motion is
\be
\phi(u,v) = f(u) - f(-1/v)
\ee
To describe the geodesic shown in Fig.\ \ref{fig:AdS2geodesics} we take the function $f(u)$ to be well-localized
with compact support around $u = 1$.  Then the ingoing part of the field
\be
\phi_{\rm in}(u) = f(u)
\ee
is a wavepacket that tracks the ingoing part of the geodesic, while the outgoing part of the field
\be
\phi_{\rm out}(v) = - f(-1/v)
\ee
tracks the outgoing part of the geodesic.  Note that the support of $\phi_{\rm in}$ begins on the right boundary and extends smoothly
across the future horizon into the interior of the black hole.

Next we consider the more general situation of a massive field in AdS${}_2$.  In this case particle geodesics are
{\sffamily S}-shaped curves which oscillate back and forth about the center of AdS.  As shown in appendix \ref{appendix:geodesics},
in Kruskal coordinates such a geodesic is given by
\bea
\label{AdS2geo}
&& u(\tau) = {\sin {1 \over 2} \left({\tau \over R} + \chi\right) \over \cos {1 \over 2} \left({\tau \over R} - \chi\right)} \\[3pt]
\nonumber
&& v(\tau) = {\sin {1 \over 2} \left({\tau \over R} - \chi\right) \over \cos {1 \over 2} \left({\tau \over R} + \chi\right)}
\eea
Here $\tau$ is proper time, $R$ is the AdS radius, and $\chi$ is related to the energy of the geodesic by $E = m \tan \chi$.  The geodesic
emerges from the past horizon at $\tau / R = - \chi$, reaches a maximum radius at $\tau = 0$, and enters the future horizon at $\tau / R = \chi$.

To construct a wavepacket that follows such a geodesic we make the ansatz
\be
\phi(u,v) = u^{-i \omega R} e^{i S(x)}
\ee
This describes a state with energy $\omega$, where the combination $x = uv$ is invariant under Rindler time translation.  We expect to
recover the geodesic (\ref{AdS2geo}) in a geometric optics limit.  Thus we consider
$\omega R \rightarrow \infty$, $m^2 R^2 = \Delta (\Delta - 1) \rightarrow \infty$ with $\omega / m \approx E / m = \tan \chi$ fixed.
That is, we take the geometric optics limit while holding the geometry of the geodesic fixed.
In this limit we can make a WKB approximation since
\be
S'(x) \sim \omega R \sim \Delta \rightarrow \infty
\ee
and
\be
\label{WKBcondition}
\big(S^\prime(x)\big)^2 \gg S^{\prime\prime}(x)
\ee
The WKB approximation turns the wave equation
\be
x(1+x)^2 {d^2 \over dx^2} e^{iS} + (1 - i \omega R) (1+x)^2 {d \over dx} e^{iS} + m^2 R^2 e^{iS} = 0
\ee
into the first-order equation
\be
{dS \over dx} = {\omega R \over 2 x} \pm \sqrt{{\omega^2 R^2 \over 4 x^2} + {\Delta^2 \over x(1+x)^2}}
\ee
The $+$ solution has the near-horizon ($x \rightarrow 0$) behavior
\be
S(x) \sim {\rm const.} \quad \Rightarrow \quad \phi(u,v) \sim u^{-i \omega R}
\ee
and describes an ingoing wave.  Likewise the $-$ solution has the near-horizon
behavior
\be
S(x) \sim \omega R \log x \quad \Rightarrow \quad \phi(u,v) \sim v^{i \omega R}
\ee
and describes an outgoing wave.  Note that there is a WKB turning point at
$x \approx - \tan^2 (\chi / 2)$ which matches the maximum radius of the geodesic (\ref{AdS2geo}).

To build a wavepacket we make a superposition of ingoing WKB waves,\footnote{Similar wavepackets were constructed in
\cite{PhysRevD.11.1404}.}
\be
\label{WKBsuperposition}
\phi_{\rm in}(u,v) = \int_{-\infty}^\infty d\omega \, a_\omega e^{-i \omega R \log u} e^{i S_{\rm in}(x)}
\ee
For the wavepacket to approximate the infalling part of the geodesic (\ref{AdS2geo}) the amplitudes
$a_\omega$ should be sharply peaked at the energy of the geodesic, that is at $\omega = E$.  The phases
of $a_\omega$ so far are arbitrary and can be absorbed into the phases of the WKB modes, so with no loss of
generality we can take the $a_\omega$ to be real and positive.

We evaluate the integral (\ref{WKBsuperposition}) in a stationary-phase approximation.  Varying with respect to
$\omega$ in the exponent, and requiring that the phase be stationary at $\omega = E$, leads to the condition
\be
\label{SaddlePoint}
\log u - \log \tan {\chi \over 2} = \int_{-\tan^2 (\chi/2)}^x {dx' \over 2 x'} \left(1 - {\tan \chi \over \sqrt{\tan^2 \chi + {4 x' \over (1+x')^2}}}\right)
\ee
Here we have fixed the phases of the WKB modes so there is constructive interference at the turning point.  That is, the stationary-phase
condition is satisfied at
\be
u = \tan (\chi/2) \qquad x = - \tan^2(\chi/2)
\ee
Evaluating the integral, (\ref{SaddlePoint}) is equivalent to
\be
u^2 = 1 + {2(x - 1) \over \tan \chi \sqrt{\tan^2 \chi \, (1 + x)^2 + 4 x} + \tan^2 \chi \, (1 + x) + 2}
\ee
This is satisfied on (\ref{AdS2geo}), so the peak of the wavepacket we have constructed moves along the desired geodesic.

The geodesic we have considered is not the most general one, since it reaches its maximum
radius at Rindler time $t = 0$.  We can find the most general geodesic by acting
with a time translation, $t \rightarrow t + t_0$.  This acts on the amplitudes by
\be
\label{shift}
a_\omega \rightarrow a_\omega e^{i \omega t_0}
\ee
The resulting geodesic has its turning point at time $t_0$, where it reaches its
maximum Rindler radius $r_0 = R/\cos \chi$.  Note that the turning point is always outside the horizon.  One can check that the stationary phase condition (\ref{SaddlePoint}) changes
appropriately under (\ref{shift}).

We have constructed wavepackets as solutions to the classical bulk equations of motion, but it is straightforward to extend
these results to the quantum theory.  In the quantum theory we could construct a coherent state $\vert \psi \rangle$ in the CFT
such that $\langle \psi \vert a_\omega \vert \psi \rangle$
is sharply localized about $\omega = E$ and has the appropriate phases.  Here
\be
a_\omega = {1 \over (2R)^\Delta} \int_{-\infty}^\infty {dt \over 2 \pi} e^{i \omega t} {\cal O}(t)
\ee
is a CFT operator modeled on (\ref{a_omega}).  Then the corresponding expectation value
\be
\langle \psi \vert \phi_{\rm in} \vert \psi \rangle
\ee
will reproduce the classical wavepacket we constructed.

This shows that, in a WKB approximation, the CFT is capable of describing a semiclassical wavepacket that
falls through the future horizon.\footnote{To some extent this follows from section \ref{sect:near}.  These
wavepackets are well-localized on the right future horizon, so by the results of section \ref{sect:near} we are guaranteed
that $\phi_{\rm in}$ accurately describes the full field in the interior.}  A key observation is that the outgoing part of the field -- which is challenging to describe in the CFT -- is simply not required to describe an infalling geodesic.  Although our formulas refer to AdS${}_2$, the wavepacket
construction is quite general and should apply to any black hole.  One simply makes a WKB approximation in the effective
potential (\ref{Veff}).  But note that in this potential, for fixed but large $\omega$ the condition for validity of the WKB approximation
(\ref{WKBcondition}) breaks down near the singularity at $r = 0$.

It would be interesting to study corrections to these infalling geodesics, arising from large but finite $N$ or from wavepackets
with finite frequency.  But we are starting from a collection of well-defined operators in the CFT.  So we expect such corrections to be
calculable and small, governed for example by the rules of the $1/N$ expansion.

\section{The black hole interior\label{sect:interior}}

So far we have argued that in the presence of a horizon a field can be decomposed into ingoing and outgoing pieces.  The ingoing
piece can be represented as an operator in a single CFT and is capable of describing semiclassical wavepackets falling into the black hole.
But one might still ask about reconstructing the full field (not just the ingoing piece) inside the horizon.  Here we explore the extent to which
this is possible, building on approaches developed in the literature, in three distinct contexts: eternal black holes, stable black holes
formed from collapse, and evaporating black holes.

\subsection{Eternal black holes\label{subsec:eternal}}

\begin{figure}
\center{\includegraphics[height=5.5cm]{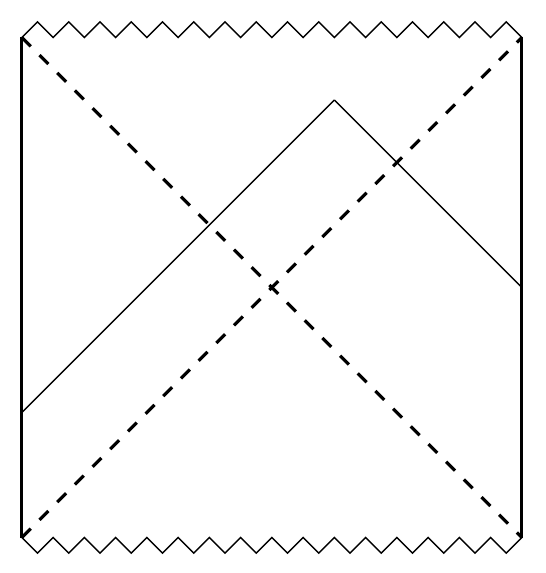} \hspace{3cm} \raisebox{-2.5mm}{\includegraphics[height=6cm]{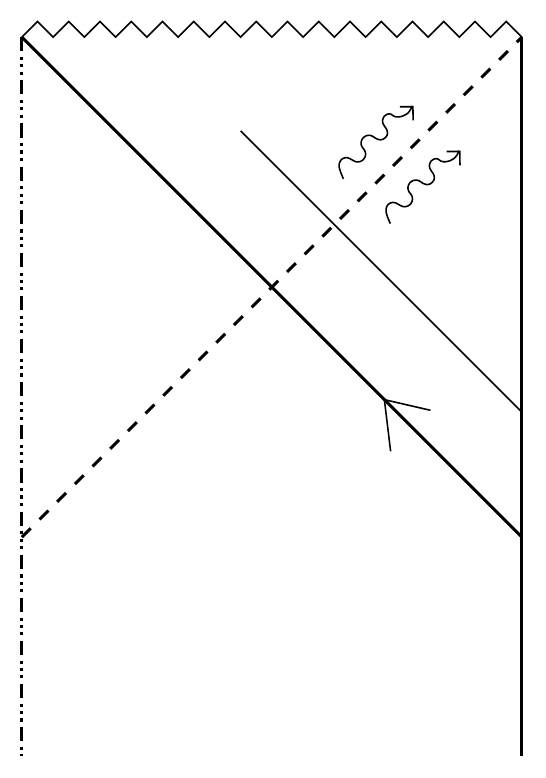}}}
\caption{On the left, an eternal black hole with two asymptotic regions.  The field in the interior is a sum of ingoing pieces from the left
and right boundaries.  On the right, a stable black hole formed by collapse.  The field in the interior is a sum of an ingoing
piece from the boundary and an outgoing piece which can be recovered from entanglement.\label{fig:EternalStable}}
\end{figure}

The simplest situation to consider is the eternal or two-sided geometry shown in Fig.\ \ref{fig:EternalStable},
provided one has access to both copies of the CFT.  In this case one can construct a field which is infalling from the right $\phi_{\rm in}^R$ and another
which is infalling from the left $\phi_{\rm in}^L$.  Each of these infalling fields can be extended to the interior,\footnote{the future interior, meaning
region II of the Penrose diagram} where one can form the superposition
\be
\phi_{\rm interior} = \phi_{\rm in}^L + \phi_{\rm in}^R
\ee
This gives the full field in the black hole interior.\footnote{This expression for the field in the interior was developed in
\cite{PhysRevD.11.1404}, where $\phi_{\rm in,out}$ were called $\phi^{+,-}$.  It was used in AdS/CFT in \cite{Papadodimas:2012aq}.}
The argument is simply that, as we will show, $\phi_{\rm interior}$ agrees with the full field $\phi$
on both the left and right parts of the future horizon.  But the future horizon provides a Cauchy surface for the black hole interior, and since it's a null
Cauchy surface the value of the field is sufficient initial data for the wave equation.
This means that $\phi_{\rm interior}$ and $\phi$ agree everywhere in the black hole interior.

Making $\phi_{\rm interior}$ and $\phi$ agree everywhere on the future horizon requires a careful treatment
of zero modes.  Recall that in section \ref{sect:near} we used an $i \epsilon$ prescription such that
\setlength\parskip{0pt}
\begin{itemize}
\item
on the right future horizon $\phi_{\rm in}^R$ agrees with the full field
\item
on the left future horizon $\phi_{\rm in}^R$ is given by the zero mode
\end{itemize}
When defining $\phi_{\rm in}^L$, we should use an $i \epsilon$ prescription such that
\begin{itemize}
\item
on the right future horizon $\phi_{\rm in}^L$ vanishes
\item
on the left future horizon $\phi_{\rm in}^L$ gives the full field minus the zero mode
\end{itemize}
This avoids double-counting the zero mode,\footnote{It also requires that the zero modes of the left and right CFT's be
identified.  This is a consistency condition for gluing two Rindler wedges together into a connected spacetime.} and with these prescriptions $\phi_{\rm interior}$ will agree with $\phi$ everywhere on the future horizon.  We study this representation of the field in more detail in appendix \ref{appendix:interior}.
\setlength{\parskip}{10pt}

\subsection{Stable black holes\label{subsec:stable}}

Next we consider the more complicated situation of a stable black hole in AdS which is formed from collapse.
Such black holes, illustrated in Fig.\ \ref{fig:EternalStable}, exist in AdS${}_5 \times S^5$
for Schwarzschild radii $R_S > R / N^{2/17}$ \cite{Horowitz:1999uv}.  In these one-sided geometries one can represent the full field in the interior in terms of a single CFT, using the construction of mirror operators developed by Papadodimas and Raju \cite{Papadodimas:2012aq,Papadodimas:2013jku,Papadodimas:2015jra}.  It is useful to view their construction in the following way.  From the bulk
perspective a smooth horizon requires an entangled state, in which supergravity degrees of freedom outside the horizon are pairwise maximally-entangled with supergravity degrees of freedom inside the horizon.  We know how to represent the outside degrees of freedom using the CFT.  We can then use the pairwise entanglement to identify corresponding
degrees of freedom in the interior.  These have a bulk interpretation as supergravity excitations inside the horizon.

In more detail, recall from (\ref{NearHorizon2}) that the outgoing modes have
the near-horizon behavior $\phi_\omega^{\rm out} \sim (-v)^{i \beta \omega / 2 \pi}$ as $v \rightarrow 0^-$.  To extend the mode across the horizon we need a prescription
for continuing past the branch point at $v = 0$.  A positive-frequency Kruskal mode\footnote{Positive frequency in the sense that it multiplies an annihilation operator in the mode expansion of the field.}
is defined by analytically continuing through the lower half of the complex $v$ plane, to obtain
\be
\phi^{\rm out,+}_\omega \sim \left\lbrace
\begin{array}{ll}
e^{-\beta \omega / 2} \, v^{i \beta \omega / 2 \pi} & \qquad \hbox{\rm as $v \rightarrow 0^+$} \\
(-v)^{i \beta \omega / 2 \pi} & \qquad \hbox{\rm as $v \rightarrow 0^-$}
\end{array}\right.
\ee
In the near-horizon region, this choice of positive frequency identifies the Kruskal vacuum for the outgoing modes -- which is locally equivalent to the Minkowski vacuum -- as a thermofield entangled
state \cite{Israel:1976ur}, where the entanglement is between degrees of freedom inside and outside the horizon.\footnote{To clarify the notation, this formula only refers to outgoing modes.
On the left we have the Kruskal vacuum for the outgoing modes.  On the right we decompose it into pieces of the outgoing modes which are supported
inside the horizon (i.e.\ at $v > 0$) $\vert \psi_i^{\rm in} \rangle$ and pieces which are supported outside the horizon (i.e.\ at $v < 0$) $\vert \psi_i^{\rm out} \rangle$.}
\be
\label{KruskalVacuum}
\vert 0 \rangle_{\rm Kruskal}^{\rm out} = {1 \over Z} \sum_i e^{-\beta E_i / 2} \vert \psi_i^{\rm in} \rangle \otimes \vert \psi_i^{\rm out} \rangle
\ee
Note that we are only considering the outgoing modes, for which $u$ is a time coordinate and $v = 0$ is an entangling surface.  The ingoing modes
are not entangled across the horizon since their modes are analytic.  But for now we will ignore the ingoing modes, since we already know how to represent them in the CFT.

Turning to the CFT, there should be a factor in the CFT Hilbert space which represents supergravity degrees of freedom outside the black hole.
Moreover the CFT state which represents the black hole should have the same entanglement structure as (\ref{KruskalVacuum}).
Given such an entangled state, following Papadodimas and Raju \cite{Papadodimas:2012aq}, to any operator on the outside Hilbert space
\be
{\cal O} = \sum_{ij} \omega_{ij} \vert \psi_i^{\rm out} \rangle \langle \psi_j^{\rm out} \vert
\ee
one can associate a mirror operator that acts on the inside Hilbert space 
\be
\widetilde{\cal O} = \sum_{ij} \omega_{ij}^* \vert \psi_i^{\rm in} \rangle \langle \psi_j^{\rm in} \vert
\ee
Since we know how to represent supergravity fields outside the black hole as operators in the CFT,
the mirror map can be applied to write supergravity fields in the interior.
Note however that the construction of mirror operators is sensitively dependent on the details of the entangled state.\footnote{For example (5.7) in \cite{Papadodimas:2012aq} must be maximally entangled
for the mirror construction to give local operators.}
In particular the mirror operators do not satisfy the ETH ansatz (\ref{ETH}), and they will only represent local operators in the interior provided one starts from a state with the specific pattern of pairwise entanglement
implied by supergravity.
This issue has been discussed in \cite{Harlow:2014yoa}.
Thus the interpretation of mirror operators as representing local degrees of freedom inside the horizon is based on
having supergravity-like entanglement across the horizon.

As an alternative to the PR construction, one could attempt to represent degrees of freedom in the
interior by evolving them backwards in time to before the black hole formed \cite{Roy:2015pga}.  For outgoing
degrees of freedom in the interior this would mean evolving backwards in time across the infalling matter,
bouncing off the left side of the Penrose diagram, and eventually reaching the exterior of the black hole.  In principle this
leads to a representation of the outgoing field in terms of a CFT operator.  However in tracing backwards it is unlikely that one can ignore
interactions with the infalling matter \cite{Almheiri:2013hfa}.  As in the PR construction, this would make the resulting CFT operator very sensitive to the microstate of the matter
which is falling in to form the black hole.  But let's imagine that we are able to evolve across the infalling shell and represent
an outgoing degree of freedom in the interior.  To check if our answer is correct we could ask whether, in the state of the CFT that represents the black hole, this degree of freedom
is maximally entangled with its expected outside partner.  This is
exactly the criterion used in the PR construction, and since maximal entanglement is monogamous it would imply that the operator
we found agrees with the PR construction.

\subsection{Evaporating black holes\label{subsect:evaporate}}

Finally we consider black holes in AdS which are formed from collapse and subsequently evaporate.  In the usual
't Hooft limit such black holes do not exist.  But as we review in appendix \ref{appendix:small}, there is a range of parameters
$N$, $\lambda$ and a range of black hole masses for which the Schwarzschild radius satisfies \cite{Horowitz:1999uv}
\be
\ell_P < \ell_s < R_S < R / N^{2/17} < R
\ee
Such small black holes are unstable and evaporate, much like black holes in flat space.

We want to ask whether an evaporating black hole has a semiclassical interior. By this we mean: are there suitable operators
in the CFT
whose correlation functions are in good agreement with the predictions of bulk effective field theory for correlators of local operators inside the
horizon.  We could attempt to build such operators using the PR construction reviewed in the previous section.  But the construction of mirror operators depends on the precise form of the entangled state.  The
pattern of trans-horizon entanglement predicted by supergravity
is plausible up to the Page time,
but past the Page time a Hawking particle that is emitted will predominantly be entangled with distant earlier Hawking radiation.
Thus the pattern of entanglement across the horizon required by local field theory is lost \cite{Braunstein:2009my}, which is the basis for the firewall proposal \cite{Almheiri:2012rt}. After the Page time there is still entanglement across the horizon,\footnote{given by the Page
curve \cite{Page:1993df,Page:1993wv}} so we can still
apply the PR construction.  But the mirror operators that it gives will not represent local degrees of
freedom in the interior.

This means that -- even using entanglement and state-dependent operators -- we are not able to represent the full bulk field
in the interior in terms of the CFT.  This suggests that the interior geometry changes at the Page time.
But semiclassical gravity would assign the black hole a well-defined interior geometry even after the
Page time: for instance geodesics approaching the horizon can be continued inside.

Since we trust the CFT it seems the gravity description must be modified.  It could be that a firewall forms, but we would
like to suggest an alternative.  The difficulty we encountered was in the CFT description of outgoing modes inside the horizon
of an old black hole.  But for ingoing modes there is no problem, and as in section \ref{sect:wavepackets} there's no difficulty describing an
infalling wavepacket in the CFT: one simply has to construct an ingoing smearing function using the evaporating geometry.\footnote{
The ingoing smearing functions have support which extends to the infinite past on the boundary.  But with the $i \epsilon$ prescription we adopted the smearing function
decays exponentially in the past, which means the ingoing field is not very sensitive to the process by which the black hole was formed.}

So the classical gravity description was not completely wrong. One can extend geodesics inside the horizon, in the sense that we can
describe wavepackets in the CFT that track geodesics in the interior exactly as one would expect for particles falling through the horizon of a classical black hole. In this sense the CFT can describe the infalling object shown in Fig.\ \ref{fig:evaporate}.  It is important to note that this can only be done in the ray or geometric optics approximation. The existence of an interior geometry after the Page time is not seen by recovering local bulk correlation functions from the CFT, as can be done before the Page time.  Instead the CFT gives us a
more bare-bones structure, in which we recover geodesics from the ray approximation for infalling wave packets.

\begin{figure}
\center{\includegraphics[width=9cm]{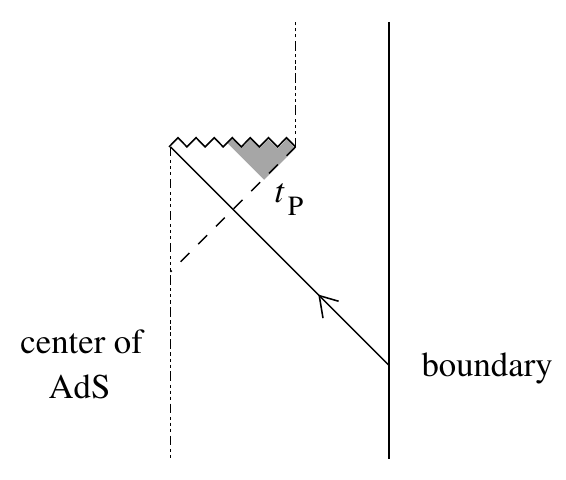}}
\caption{In an evaporating black hole trans-horizon entanglement is lost at the Page time $t_{\rm P}$, so local outgoing
degrees of freedom do not exist in the shaded region.
\label{fig:breakdown}}
\end{figure}

Thus the CFT leads us to an asymmetric picture of the interior of an old black hole, illustrated in Fig.\ \ref{fig:breakdown}.
According to the outgoing modes, which are responsible for Hawking evaporation, a local geometry exists in the interior
only as long as the interior has a specific pattern of entanglement with the outside.  But according to the ingoing modes,
which are capable of describing infalling classical observers, a well-defined classical interior geometry exists at all
times.

{\em In this sense AdS/CFT reconciles unitarity of the evaporation process with the classical geometry seen by an infalling observer.}

\section{Conclusions\label{sect:conclusions}}

In this paper we used the construction of local bulk observables to gain insight into the black hole
interior.  We found that in a one-sided geometry the CFT makes a sharp distinction between ingoing and outgoing fields.
Ingoing fields can be represented as conventional smeared operators in the CFT and can be used to describe
infalling geodesics.  Outside the horizon the outgoing fields can be represented as conventional CFT operators.  In
the interior
they can only be accessed using entanglement.  But past the Page time the trans-horizon entanglement no longer agrees
with
supergravity expectations, which means there is no CFT representation of local right-moving degrees of freedom in the interior.
It seems the existence of a local internal geometry depends on entanglement, as suggested in \cite{VanRaamsdonk:2010pw}.

It's tempting to speculate that this partial breakdown of locality provides a mechanism for transporting information out of the black hole
interior.  Up to the Page time outgoing modes in the interior can be described via their pairwise entanglement with supergravity degrees of freedom
outside the black hole.  Note that these outgoing modes have propagated through the infalling matter, so their quantum state should
be sensitive to the details of the matter that fell in to make the black hole.  Starting around the Page time these outgoing degrees of
freedom no longer have a local description.  There are still outgoing degrees of freedom inside the black hole -- the black hole still has entropy,
and entanglement across the horizon is given by the Page curve -- but these degrees of freedom no longer behave locally.  This opens the
possibility for them to transport information about the state of the infalling matter out to a stretched horizon where locality is restored.

There are many directions in which this new picture of the black hole interior could be further developed and understood.
In this paper we only considered free scalar fields.  It would be interesting to extend the results to fields with spin and
understand the subtleties associated with gauge invariance \cite{Harlow:2015lma,Guica:2015zpf}.
Perhaps more importantly, it would be interesting to extend the construction beyond the free-field limit.  
In this paper we have shown that the CFT provides a description of infalling geodesics
even after the Page time.  This is consistent with, but does not imply, the idea that an infalling observer experiences a smooth horizon.
For example the observer could carry a particle detector (or a thermometer) coupled to the field, or could be performing experiments at low energy in the observer's frame.
To what extent can such observations and experiments be described by the CFT, and do they give results that are consistent with a smooth horizon?

\bigskip
\bigskip
\goodbreak
\centerline{\bf Acknowledgements}
\noindent
We are grateful to Dorit Aharonov, Max Brodheim, Gary Gibbons and Ian Morrison for valuable discussions.  We thank the Aspen
Center for Physics and the KITP for hospitality.  At ACP this work was supported in part by National Science Foundation grant
PHY-1066293 and at KITP by NSF grant PHY11-25915.
DK is supported by U.S.\ National Science Foundation grants PHY-1214410 and PHY-1519705 and is grateful to the Columbia University Center for Theoretical Physics for
hospitality.  GL is supported in part by the Israel Science Foundation under grant 504/13 and thanks the City University of New York for hospitality.

\appendix
\section{Geodesics in AdS${}_2$\label{appendix:geodesics}}

To obtain the massive geodesics used in section \ref{sect:wavepackets} it's convenient to represent AdS${}_2$
as a hyperboloid
\be
-(X^0)^2 - (X^1)^2 + (X^2)^2 = - R^2
\ee
inside ${\mathbb R}^{2,1}$ with metric
\be
ds^2 = -(dX^0)^2 - (dX^1)^2 + (dX^2)^2
\ee
The obvious timelike geodesic winds around the waist of the hyperboloid.
\bea
\nonumber
&& X^0 = R \cos (\tau / R) \\
\label{AtRest}
&& X^1 = R \sin (\tau / R) \\
\nonumber
&& X^2 = 0
\eea
A more general geodesic can be obtained by acting with a Lorentz boost.
\be
\label{boost}
\left(\begin{array}{c} X^0 \\ X^1 \\ X^2 \end{array}\right) =
\left(\begin{array}{ccc} \cosh \phi & 0 & \sinh \phi \\ 0 & 1 & 0 \\ \sinh \phi & 0 & \cosh \phi \end{array}\right)
\left(\begin{array}{c} R \cos (\tau/R) \\ R \sin (\tau/R) \\ 0 \end{array}\right)
\ee
Introducing Rindler coordinates via
\be
X^0 = r \qquad X^1 = \sqrt{r^2 - R^2} \, \sinh(t/R) \qquad X^2 = \sqrt{r^2 - R^2} \, \cosh(t/R)
\ee
the geodesic becomes
\bea
\label{RindlerGeodesic}
&&t(\tau) = R \tanh^{-1}\big({\tan(\tau / R) / \sinh \phi}\big) \\
\nonumber
&&r(\tau) = R \cos(\tau / R) \cosh \phi
\eea
In terms of the Kruskal coordinates introduced in (\ref{KruskalCoords}), and setting $\cosh \phi = 1 / \cos \chi$, this
gives (\ref{AdS2geo}).

In fact $\chi$ parametrizes the energy of the geodesic.  To see this note that for a particle of mass $m$ a metric
of the form (\ref{StaticMetric}) gives rise to a conserved energy $E = m f(r) {dt \over d\tau}$.
Evaluating this on (\ref{RindlerGeodesic}) gives $E = m \sinh \phi = m \tan \chi$.

\section{Small unstable black holes in AdS\label{appendix:small}}

In section \ref{subsect:evaporate} we considered black holes in AdS which are formed from collapse and
subsequently evaporate.  Such black holes can be described in terms of the CFT, but one has to work in
a non-'t Hooft limit.  Here we review the construction, following the work of Horowitz \cite{Horowitz:1999uv}.

For definiteness we consider four-dimensional ${\cal N} = 4$ supersymmetric $SU(N)$ Yang-Mills with 't Hooft
coupling $\lambda = g^2_{\rm YM} N$, dual to string theory on AdS${}_5 \times S^5$ with string coupling and AdS
radius
\be
g_s = g^2_{\rm YM} = \lambda / N \qquad\quad R = \lambda^{1/4} \ell_s
\ee
The 10-dimensional Planck length is
\be
\ell_P = g_s^{1/4} \ell_s = R / N^{1/4}
\ee
The thermal phases of interest are
\begin{itemize}
\item
a 10-dimensional supergravity gas with microcanonical entropy
\be
S_{\rm gas} \sim (R E)^{9/10}
\ee
\item
a stringy Hagedorn phase with entropy
\be
S_{\rm Hagedorn} \sim E \ell_s
\ee
\item
a 10-dimensional black hole which is small in the sense that the Schwarzschild radius $R_S < R$.  The energy and
entropy are
\be
E_{\rm bh} \sim R_S^7/\ell_P^8 \qquad\quad S_{\rm bh} \sim R_S^8 / \ell_P^8
\ee
\end{itemize}

We're interested in black holes that behave much as in flat space, that are formed from collapse and subsequently
evaporate to a gas of gravitons.  To achieve this in AdS/CFT we want
\begin{itemize}
\item
$\lambda > 1$ and $N > 1$ so the AdS radius is large compared to the string and Planck lengths:
$R > \ell_s$ and $R > \ell_P$
\item
$N > \lambda$ so the string theory is weakly-coupled: $g_s < 1$ and $\ell_P < \ell_s$
\item
a Schwarzschild radius which is large compared to the string and Planck lengths, so the black hole
behaves semiclassically
\item
a Schwarzschild radius which satisfies $R_S < R / N^{2/17}$, so the black hole has less entropy than
a graviton gas of the same energy: $S_{\rm bh} < S_{\rm gas}$.  Such a black hole is unstable
and will evaporate.
\end{itemize}

\begin{figure}
\center{\hspace*{1cm} \includegraphics[width=9cm]{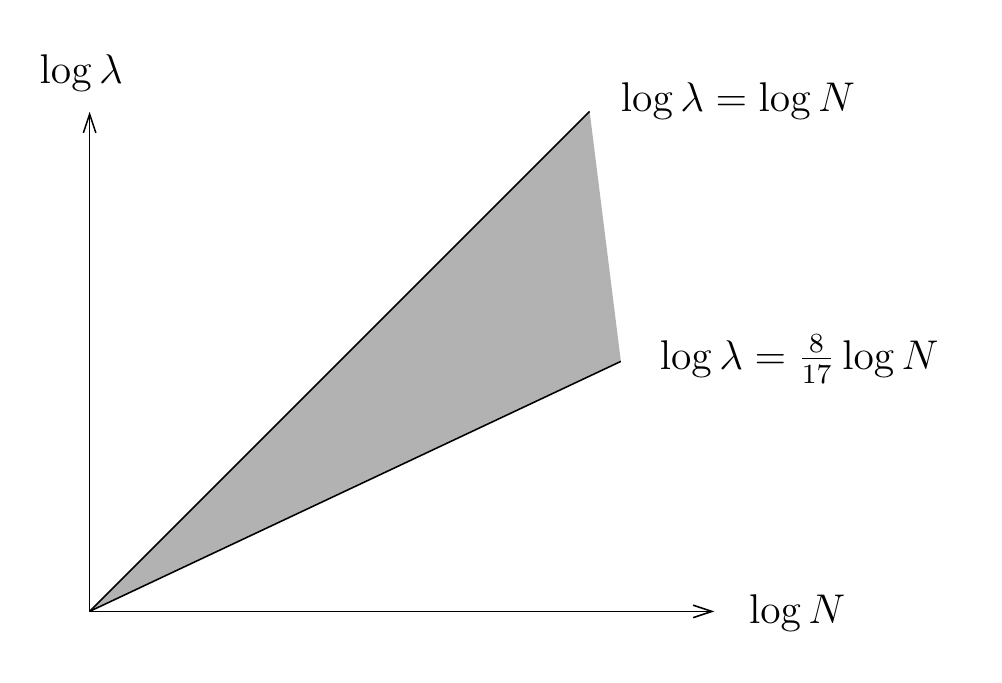}}
\caption{The range of Yang-Mills parameters for which small unstable black holes exist in AdS${}_5$.
\label{fig:range}}
\end{figure}

To summarize we're interested in the range of parameters shown in Fig.~\ref{fig:range},
\be
N \gg \lambda \gg 1 \quad {\rm and} \quad
\lambda^{1/4} \gg N^{2/17}
\ee
Given such parameters there's a range of Schwarzschild radii for which
\be
\ell_P \ll \ell_s \ll R_S \ll R / N^{2/17} \ll R
\ee
In this range we have
\be
S_{\rm Hagedorn} \ll S_{\rm bh} \ll S_{\rm gas}
\ee
and the black hole evaporates as though it were in flat space.  Note that such black holes do not exist in the
usual strongly-coupled 't Hooft limit, where $N \rightarrow \infty$ with $\lambda \gg 1$ fixed.

\section{Fields inside an eternal black hole\label{appendix:interior}}

In section \ref{subsec:eternal} we gave a prescription for defining the field in the future interior of an eternal black
hole as a sum $\phi_{\rm in}^L + \phi_{\rm in}^R$ of ingoing fields from the left and right boundaries.  Here we
explore this prescription in more detail and show that it is compatible with other expressions in the literature.

\begin{figure}
\center{\includegraphics{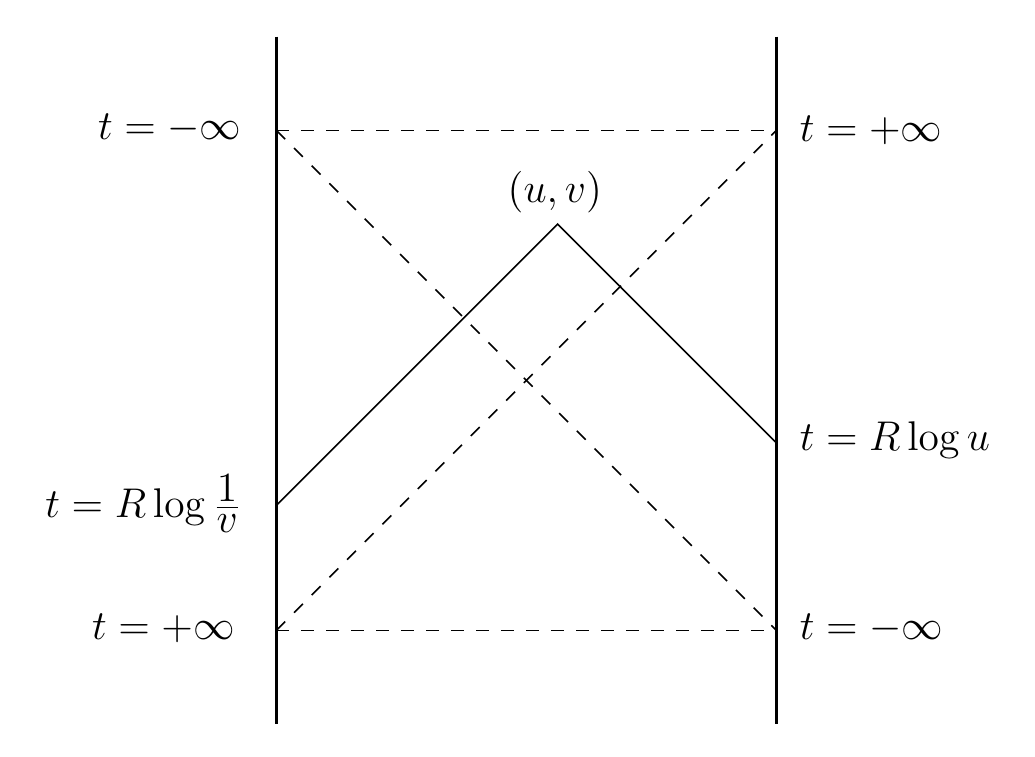}}
\caption{The smearing function for a point in the interior is non-analytic when past-directed null rays from the bulk point
hit the boundary.\label{fig:InteriorSmear}}
\end{figure}

We work in AdS${}_2$ in the Rindler patch and consider fields with $\Delta = 1$ and $\Delta = 2$.  Expressions for $\phi_{\rm in}^R$ are given in (\ref{DeltaEq1phiin}) and (\ref{DeltaEq2phiin}), but we should be more explicit about the form of $\phi_{\rm in}^L$.  With the $i \epsilon$ prescription described in section \ref{subsec:eternal} we find that for $\Delta = 1$
\be
\phi_{\rm in}^L = - {1 \over 2 R^2} \int\limits_{R \log (1/v)}^\infty \!\! dt \, {\cal O}^L(t)
\ee
and for $\Delta = 2$
\bea
\label{DeltaEq2phiinL}
\phi_{\rm in}^L & = & - {3 \over 2 R^3} \int_{-\infty}^{R \log (1/v)} dt \, {1 \over 1 + uv} \, v e^{t/R} \, {\cal O}^L(t) \\
\nonumber
& & + {3 \over 2 R^3} \int_{R \log (1/v)}^\infty dt \, {1 \over 1 + uv} \left(- 1 + uv - u e^{-t/R}\right) {\cal O}^L(t)
\eea
Here ${\cal O}^L$ is an operator in the left CFT.  Time runs up on the right boundary and
down on the left, as shown in Fig.\ \ref{fig:InteriorSmear}.  A heuristic way to obtain these results is to (i) start with
$\phi_{\rm in}^R$, (ii) replace $R \log u \rightarrow R \log (1/v)$ in the limits of integration, and (iii) change the
sign of the constant term present in the smearing function at late times.  Likewise to obtain $\phi_{\rm out}^L$ one starts with
$\phi_{\rm out}^R$, replaces $R \log (-1/v) \rightarrow R \log (-u)$, and flips the sign of the constant term at late times.

For $\Delta = 1$ the result for
\be
\phi_{\rm interior} = \phi_{\rm in}^L + \phi_{\rm in}^R
\ee
agrees with (39) in \cite{Hamilton:2005ju}.  But for $\Delta = 2$ the two expressions are different, and it is not obvious
that they will agree inside correlation functions.
We will show that the expressions for $\Delta = 2$ are in fact compatible by transforming to global coordinates and explaining
in what sense they agree.

Kruskal and global coordinates are related by
\be
u = \tan {\tau + \rho \over 2} \quad\qquad
v = \tan {\tau - \rho \over 2}
\ee
This puts the metric in the form
\bea
&& ds^2 = {R^2 \over \cos^2 \rho} \big(-d\tau^2 + d\rho^2\big) \\
\nonumber
&& -\infty < \tau < \infty \qquad - {\pi \over 2} < \rho < {\pi \over 2}
\eea
Global time $\tau$ is related to Rindler time on the left and right boundaries by
\be
\tanh ({t_L / R}) = - \sin \tau \qquad
\tanh ({t_R / R}) = + \sin \tau
\ee
Also the boundary fields in Rindler and global coordinates are related by
\be
\phi \sim {1 \over r^\Delta} \phi_0^{\rm Rindler} \sim \cos^\Delta \! \rho \, \phi_0^{\rm global}
\ee
which implies
\be
\phi_0^{\rm Rindler} = (R \cos \tau)^\Delta \phi_0^{\rm global}
\ee
With these ingredients it is straightforward to transform $\phi_{\rm interior}$ to global coordinates.
There is one more fact we need: in AdS${}_2$, for fields with integer dimension, the antipodal
map relates fields on the left and right boundaries by \cite{Hamilton:2005ju}
\be
\phi_0^{\rm global, L}(\tau) = (-1)^\Delta \phi_0^{\rm global, R}(\tau + \pi)
\ee
This lets us rewrite $\phi_{\rm interior}$ in global coordinates purely in terms of the right boundary field.
We find
\bea
\nonumber
&&\hspace*{-5mm}\phi_{\rm interior}(\tau,\rho) = {3 \over 2} \int_{\tau - ({\pi \over 2} - \rho)}^{\tau + ({\pi \over 2} - \rho)} d\tau'
\left({1 - u v \over 1 + uv} \cos \tau' - {u \over 1 + uv} \big(1 - \sin \tau'\big)\right) \phi_0^{\rm global,R}(\tau') \\
\label{RightBdySmear}
&& \hspace{5mm} - {3 \over 2} \bigg(\int_{-\pi/2}^{\tau - ({\pi \over 2} - \rho)} + \int_{\tau + ({\pi \over 2} - \rho)}^{3\pi/2}\bigg)d\tau' \,
{v \over 1 + uv} \big(1 + \sin \tau'\big) \, \phi_0^{\rm global,R}(\tau')
\eea
(In this expression $u,v$ are the Kruskal coordinates of the bulk point.)

At this point it's important to recognize that smearing functions are not unique.  In global coordinates, for a field
of integer conformal dimension, the boundary field is $2\pi$ periodic in global time but the Fourier components
with frequencies $-\Delta+1,\ldots,\Delta-1$ are absent \cite{Hamilton:2005ju}.  For $\Delta = 2$
this means we're free to add terms to the smearing function with time dependence $1,e^{i\tau},e^{-i\tau}$.
We can use this freedom to eliminate the second line of (\ref{RightBdySmear}), leaving\footnote{Note that the integral is over spacelike-separated points on the right boundary.}
\be
\phi_{\rm interior}(\tau,\rho) = {3 \over 2} \int_{\tau - ({\pi \over 2} - \rho)}^{\tau + ({\pi \over 2} - \rho)} d\tau' \,
{\cos (\tau - \tau') - \sin \rho \over \cos \rho} \, \phi_0^{\rm global,R}(\tau')
\ee
in agreement with the global smearing function obtained in \cite{Hamilton:2005ju}.  This is
another example of the non-uniqueness of smearing functions that was studied in \cite{Almheiri:2014lwa,Mintun:2015qda}.

As a further check we used the expressions (\ref{DeltaEq2phiin}), (\ref{DeltaEq2phiinL}) for $\phi_{\rm in}^R$ and $\phi_{\rm in}^L$ to compute a bulk-to-boundary 2-point function.  Starting from thermal correlators in the CFT we recovered, as expected, a
bulk-to-boundary correlator in the Kruskal vacuum.


\providecommand{\href}[2]{#2}\begingroup\raggedright\endgroup

\end{document}